\documentclass[11pt]{article} 
\usepackage{geometry} % to change the page dimensions
\usepackage{amssymb}
\usepackage{amsthm}
\usepackage{amsmath}
\usepackage{color}
 \usepackage{tabu}
\usepackage{multirow}
\usepackage{array}
\usepackage{algpseudocode}
\usepackage{algorithm}
\usepackage{slashbox}
\usepackage{graphicx}
\usepackage[font=footnotesize,labelfont=bf]{caption} 
\newtheorem{theorem}{Theorem}
\newtheorem{remark}[theorem]{Remark}

\title{Mathematical modeling of local perfusion \\ in 
large distensible microvascular networks}
\author{Paola Causin\footnote{Dept of Mathematics, Universit\`a degli Studi 
di Milano, e-mail: {\tt paola.causin@unimi.it}}, Francesca Malgaroli\footnote{Dept of Mathematics, Politecnico di 
Milano, e-mail: {\tt francesca.malgaroli@polimi.it}}}

\date{}

\begin{document}
\maketitle

%\begin{frontmatter}
%
%
%\title{Mathematical modeling of local perfusion \\ in 
%large distensible microvascular networks}
%
%%% use optional labels to link authors explicitly to addresses:
%\author[Causin]{Paola Causin}
%\address[Causin]{Dipartimento di Matematica, Universit\`a degli Studi di Milano, Italy \\
%              e-mail:{\hspace{.05mm} paola.causin@unimi.it} }
%
%\author[Malgaroli]{Francesca Malgaroli}
%\address[Malgaroli]{Dipartimento di Matematica, Politecnico di Milano, Italy}

\begin{abstract}

Microvessels -blood vessels with diameter less than 200~$\mu$m-
form large, intricate networks organized into arterioles, capillaries and venules. 
In these networks, the distribution of flow and pressure drop is a highly interlaced function of single vessel resistances
and mutual vessel interactions. Since, it is often impossible to quantify all these aspects when collecting 
experimental measures, in this paper we propose 
a mathematical and computational model to study the behavior of microcirculatory networks 
subjected to different conditions. 
The network geometry, which can be derived from digitized images of experimental measures 
or constructed {\em in silico} on a computer by mathematical laws, is simplified for computational purposes into a graph of
connected straight cylinders, each one  
representing a vessel.    
The blood flow and pressure drop across the single vessel, further split into smaller elements, are related through a generalized Ohm's law featuring a conductivity parameter, function of the vessel cross section area and geometry, which undergo deformations under pressure loads. 
The membrane theory is used for the description of the deformation of vessel lumina, tailored to the structure of thick--walled arterioles
and thin--walled venules.
In addition, since venules can possibly experience negative values of transmural pressure (difference between
luminal and interstitial pressure), a buckling model 
is also included to represent vessel collapse. 
The complete model including arterioles, capillaries and venules represents a nonlinear coupled system of PDEs, which is approached numerically by finite element discretization and linearization techniques.  
As an example of application, we use the model to simulate flow in the microcirculation of the human eye retina,
a terminal system with a single inlet and outlet. 
After a phase of validation against experimental measurements of the correctness of the blood flow and pressure fields
in the network, we compute the network response to different interstitial 
pressure values. Such a study is carried out both for global and localized variations of the interstitial pressure.  
In both cases, significant redistributions of the blood flow in the network arise,     
highlighting the importance of considering the single vessel behavior along with its position and connectivity
in the network. 

\bigskip 

{\small \hspace*{2cm} Vascular Resistance;  Retinal Microcirculation ;  Mathematical Model  }

\end{abstract}

\section{Introduction}	

Microscopic blood vessels play the vital role of locally perfusing single body's organs. 
These  circulatory districts include thousands of microvessels
(diameter less than 200$~\mu$m),  
categorized as arterioles, capillaries or venules, according to their structure and function.\\
First elementary assessments of microcirculatory mechanisms on skin
or superficial organs   
date back at least to the 18$^{\rm th}$ century~\cite{Pries2003}.
At present, techniques like positron emission
tomography, magnetic resonance imaging and contrast echography
allow to study in a non--invasive manner regional blood flow in internal organs of human patients.
The information content of such measurements is, however,
far to be complete, both in baseline conditions as
well as in altered conditions. 
As a matter of fact, data such as vessel geometry, fluid-dynamics and physical parameters in physiological
conditions or in altered conditions/provocation studies often cannot  be coherently and comprehensively collected. 
These facts prevent from observing in detail    
the distribution of blood flow and pressure drop within the large and intricate microvascular networks,
 which are highly interlaced function of vessel resistances, in turn determined by
blood biophysical properties,  
vessel mechanical and geometrical features~\cite{microcircbook,Fung2013}.  
For this reason, theoretical and computational models can help reducing this gap in
knowledge.

\medskip

Mathematical models in hemodynamics are present in literature since the 1960s. Nowadays, they are well assessed tools in
 the simulation of a limited number of vessels of major size like the aorta and collaterals, 
 possibly coupled with reduced--order models for the rest of the circulatory system~(see, {\em e.g.}, 
the review works~\cite{Formaggia2009,Brunberg2009}).
In the context of microcirculation, the presence of  
an exceedingly large number ($> 10^4$) of connected vessels with complex behavior
makes the problem pretty different and requires to adopt specific techniques. 
 Typically, models of microcirculation use sets of representative segments to 
 describe a network of vessels of different size 
(for example regrouping large/small arterioles and venules, see, 
{\em e.g.},~\cite{Ye1994,Ursino1998,Arciero2013,Guidoboni2014}).
This approach maintains a low number of unknowns and
allows to explore in a simple manner different regulatory
mechanisms via phenomenological relations. 
What is lost is the spatial distribution of field variables, so that  
relevant geometrical and physical heterogeneities of the network cannot 
be represented, as well as complex internal 
interactions (see~\cite{Roy2012} for a discussion on this topic).
Spatial heterogeneity is taken into account in a different set of papers, which represent 
relatively small vessel graphs as a collection of 1D distensible tubes, 
derived from mathematical algorithms~\cite{Dawson2003, Boas2008,David2009,Causin2015a} or from
medical imaging data~\cite{Fry2013}.
While these models include the effect of the geometrical localization of each vessel
 in the network the mechanical description is absent (rigid vessels) or  
very simplified. In this latter case, phenomenological vessel compliance laws are often used,
which reproduce selected structural behaviors. 
  Moreover, it is not kept into consideration the fact that 
microcirculatory networks, characterized by low 
values of the luminal pressure, comparable to the surrounding interstitial pressure, can experience
severe reductions of the luminal cross section, till collapse,  
both in physiological conditions and -more dramatically- in presence of pathologies. 
The very few works addressing this phenomenon locate themselves at two
 opposites. Either, they study a single vessel (or very 
 small networks)  with a complex description, possibly 3D and anatomically accurate,  
 ~\cite{Lee2004,Bols2013,Ho2013,Kozlovsky2014,Heil2016,Aletti2016},  or
 they model mid-sized/large microvascular networks including a Starling/collapsible
 components~\cite{Muller2014b,Ursino1997,Ursino1998} or phenomenological variations of
 the physical parameters (see~\cite{ContarinoSIMAI2016} in a slightly different context). 
 While the first approach is unable to be applied to a large network due 
 to its huge computational cost, the second one is much more efficient but it does not consider 
 the sophisticated coupling between vessel mechanics and pressure loads.
These facts limit the spectrum of phenomena which can be analyzed.

\medskip

In this article, we propose a  
mathematical model of general microcirculatory districts 
which is capable of
dealing with large, general networks with an affordable time of resolution. 
Upon extracting the geometry from medical imaging data or from a computer--generated structure, each network 
is described as a graph of distensible tubes (vessels), representative of 
realistic structures under physiological pressures.   
A simplified fluid--structure interaction approach is considered. 
Namely, blood flow in each vessel is accounted for by a generalized Poiseuille's law, 
featuring a conductivity parameter which is function, among the others, of the area and shape of the tube cross section. 
To dispose of these latter parameters, 
we adopt thick or thin--wall structural models according to the physiological  
vessel wall thickness-to-radius ratio~\cite{Canic2006,Sriram2012}.    
In addition, a buckling model for thin--walled vessels derived from~\cite{Flaherty1972}
is included.  Buckled vessels  lose the circular--shaped cross section and assume an elliptical or dumb--bell 
configuration, causing a strong increase of vessel resistance
to flow.  The model can withstand partial or total vessel blockage, 
providing detailed information about the fluid-dynamics. 
This contrasts with simpler Starling resistor elements, which due to their
switch--like behavior, cannot represent intermediate regimes (partial patency to flow).   
Using the network geometries 
proposed in~\cite{Takahashi2009,Takahashi2014}, we simulate blood flow in 
the retinal circulation, which we already studied without including vessel compliance in~\cite{Causin2015b,Causin2016}.   
The network is composed of more than~9000 vessels, with a tunable degree of asymmetry.  
After validating of the model against experimental measures
(data from~\cite{Riva1985}), we carry out two sets of studies: 
{\em i)} we globally increase the external pressure, reaching the conditions
for buckling to occur. We observe that the luminal pressure gradually increase along all the network, till buckling, 
after which a discontinuity in the behavior takes place, with a much more marked
pressure increase and flow redistribution; {\em ii)} we locally increase the external pressure inside a spherical
region located in correspondence of a region of the post--capillary veins. This perturbation is observed to extend its effect  
till four or five vessels generations away, with important redistribution of flow and resistance.  
An interesting, noticeable, key point emerging from the above results is the 
importance of vessel localization in the network. Vessels with the same mechanical and
geometrical properties but laying in two different regions of the network display a pretty different behavior due 
to  their local pressure levels and interaction with other vessels.  

\medskip

The paper is organized as follows: in Sect.~\ref{sec:micromodel}
we present our mathematical model for microcirculatory districts. Namely,
in Subsect.~\ref{sec:bloodflow} we introduce the mathematical model 
for blood flow in a single vessel  and the generalized Ohm's law connecting flow rate and pressure drop
via the conductivity parameter. This latter is obtained in a coupled manner from the wall structure models
as discussed in Subsect.~\ref{sec:configurations} in pre--buckling conditions 
(see~\ref{sec:pre-buckl}), where a particular attention is devoted to the 
Young modulus choice, and buckled conditions
(see ~\ref{sec:thin}). In Sect.~\ref{sec:riassunto}, we report a summary of the
computation of the conductivity parameter.  
In Sect.~\ref{sec:unloaded}, we discuss the importance of using a correct
unloaded configuration, describing the numerical  technique applied to compute it
from measurements. 
In Sect.~\ref{sec:net}, we introduce the nomenclature to deal with a network
and we present the conditions to couple single vessels converging in a node.   
In Sect.~\ref{sec:solproc}, we provide a summary of the model (see~\ref{sec:model-summ}) and 
 we discuss the numerical procedure employed to 
discretized the fully coupled problem (see~\ref{sec:num-approx}). Then, in Sect.~\ref{sec:numres},
we first introduce the network geometries we will use in simulations (see~\ref{sec:geo-net})
along with the physical parameters (see~\ref{sec:phys-par})
then,  we present the results of the numerical simulations in different
test cases  (see~\ref{sec:test-cases}).
Eventually, in Sect.~\ref{sec:conclu}, we draw the conclusions,
discussing  the results along with their significance, the limitations of the model and the 
forthcoming work.

%%%%%%%%%%%%%%%%%%%  MODEL EQUATIONS %%%%%%%%%%%%%%%%%%%%%%%%%%

\section{Microcirculation model}
\label{sec:micromodel}

\subsection{Geometrical  model}
The geometry, denoted in the following as the ``measured geometry'', of the network can be 
originally derived from digitized images or can
be constructed {\em in silico} on a computer on the basis of anatomical data. In any case,  
our starting point is the 1D skeleton of the network along with its 
topological connectivity and cross sections and lengths distribution. 
Each segment of the skeleton represents a blood vessel
and can branch at nodal junctions. 
To increase the computational accuracy, we introduce further subdivisions into elements on each 
segment (see~\cite{Ho2013} for a similar approach).
 We establish on each element a local system of cylindrical coordinates and we let the $z$-axis
coincide with the element axis, arbitrarily choosing its orientation.
The vessel element is endowed of the 3D structure of a straight cylinder of axis~$z$, 
with uniform, but not necessarily circular, cross section. 
The $r$ and $\theta$ coordinates lay in the plane
of the vessel cross section (see Fig.~\ref{fig:domain}).
Elements belonging to the same vessel share homogeneous mechanical properties. 
From this geometrical model, 
we proceed by computing a reference (``unloaded'') configuration
as described in Sect.~\ref{sec:unloaded}. This latter geometry 
represents the mathematical domain of the present model. 

\subsection{Blood flow model}
\label{sec:bloodflow}

\begin{figure}[t]
\centering
\includegraphics[scale=1]{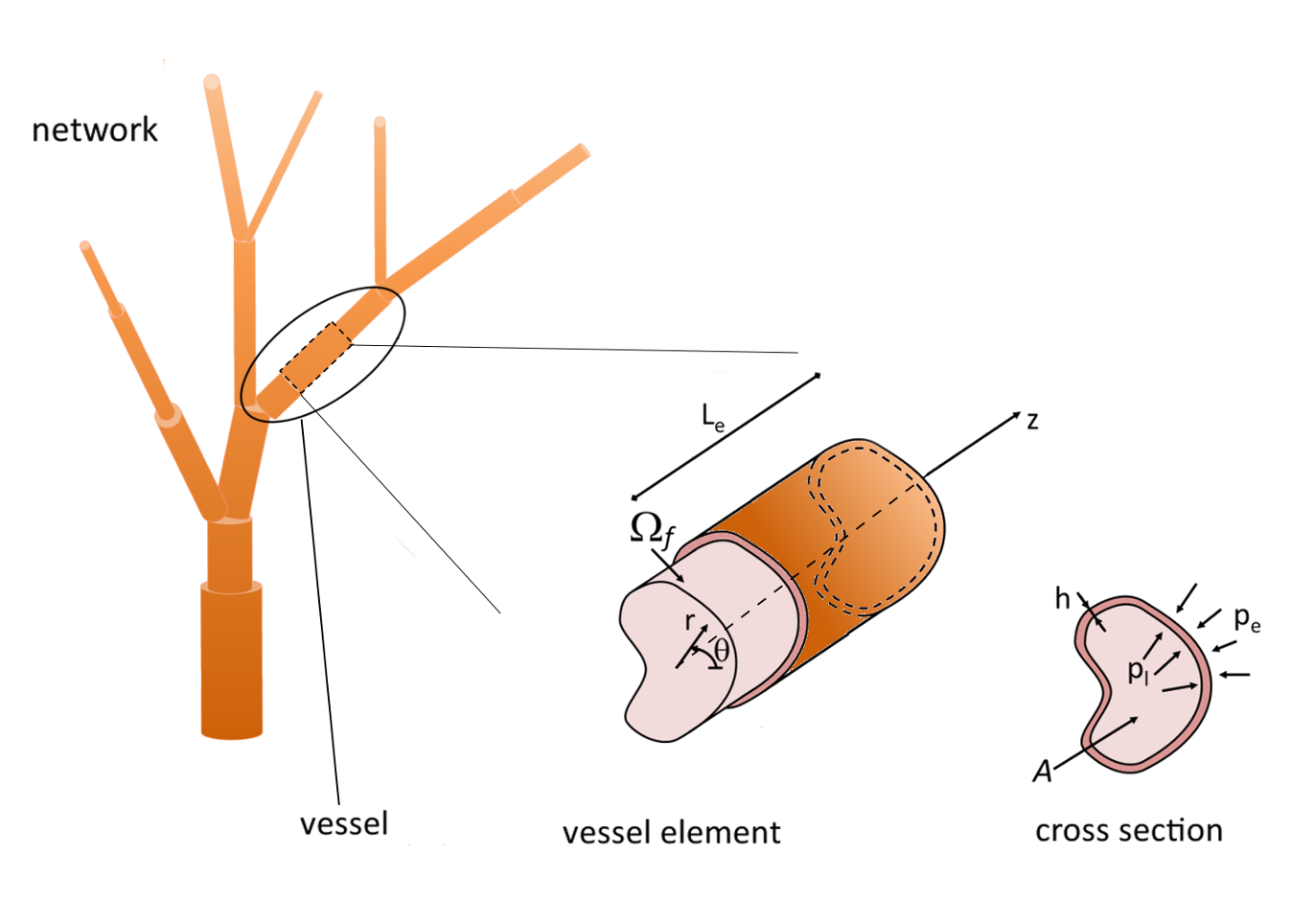}
\caption{Schematic representation of a portion of a microcirculatory network. 
Each vessel is described as a duct with straight longitudinal axis and 
is further partitioned  into a series of consecutive short elements
of arbitrary but constant cross section shape along the axis. 
The central  part of the figure depicts
one of such elements, of constant 
length~$L_e$, with highlighted   
the domain~$\Omega_f$ occupied by the blood flowing inside the luminal space 
and the wall structure. 
The right part of the figure represents the  
cross section~$A$ of the luminal space along with the thickness~$h$ of the vessel wall.
The wall is internally loaded with 
pressure $p_{\text{l}}$ from fluid actions and externally loaded with 
given interstitial pressure $p_{\text{e}}$. 
The local system of coordinates $(r,\theta,z)$  on the element is represented as well.}
\label{fig:domain}
\end{figure}

The  domain occupied by blood inside the vessel element (luminal space) is defined as (see Fig.~\ref{fig:domain})
$$
\Omega_f= A \times (0,L_e)
%\{(r,\theta,z) \in \mathbb{R}^3 \, | \, (r,\theta)\in A, z \in (0,L_e) \},
$$
where $A=(0,R(\theta)) \times (0,2\pi)$,
$R(\theta)$ being the position of the blood--wall interface and where $L_e$ is the element length. 
%and has measure given by
%$
%|A|=\int\displaylimits_{(r,\theta)\in \mathcal{D}}r\, dr\, d\theta.
%$
Blood circulation is
modeled as the steady unidirectional flow of a Newtonian incompressible fluid with dynamic viscosity~$\mu$. 

Letting~$p$ be the fluid pressure and $u$  the axial (and only) velocity component,  
the continuity and momentum balance equations  read:\\
find $p$ and $u$ such that 
\begin{equation}
\dfrac{\partial u}{\partial z} =0, \qquad  
\Delta_{r\theta} u=\dfrac{1}{\mu}\dfrac{\partial p}{\partial z}, \qquad
\dfrac{\partial p}{\partial r}=\dfrac{\partial p}{\partial \theta}=0
\qquad\qquad \text{in}\,\, \Omega_f,
\label{eq:cons1}
\end{equation} 
where $\Delta_{r\theta}(\cdot)$ is the Laplacian operator with respect to the $(r,\theta)$ coordinates.
No-slip conditions are considered on $\partial A \times (0,L_e)$.
Notice that the pressure field resulting from Eqs.~\eqref{eq:cons1}
has a constant gradient in the axial direction.
Moreover, the pressure is uniform on each section, so that,  
straightforwardly, the fluid pressure~$p_{\text{l}}$ acting on the internal surface 
of the wall structure is equal to~$p$. 

\begin{remark}
In this work, we consider a dynamic viscosity
depending, among the others, on the vessel cross section diameter (more generally speaking on the hydraulic diameter), 
as discussed in detail in Sect.~\ref{sec:phys-par},
yielding a nonlinear coupling with the geometry. This aspect is dealt with numerically by resorting to 
an iterative technique, which results
at each iteration the viscosity to be a constant, given, value, computed from quantities known from the past 
internal iteration (denoted here tout-court by $\mu$ with a slight abuse of notation).
\end{remark}

Our goal is to obtain a form of Eqs.~\eqref{eq:cons1} which is amenable to be
efficiently coupled with wall structure equations in the context of 
a large network of vessels. Introducing   
the non--dimensional variables
$ r^*={r}/{\widehat{R}}$ and $u^*=(-\mu\, \big(\widehat{R}^2 {d p}/{d z}\big)^{-1})u $, 
$\widehat{R}$ being a characteristic linear dimension of the cross section,
we write the dimensionless form of Eq.~\eqref{eq:cons1}$_2$ as~\cite{White2006}:\\
find $u^*$ such that 
\begin{equation}
-\Delta_{r^*\theta}\, u^* =1 \qquad   \qquad  \text{in}\,\, A^*, 
\label{eq:lapla}
\end{equation}
where $A^*=A/\widehat{R}^2$, and  
$u^*=0 \, \, \text{on}\,\, \partial A^*$.
We use the 
solution~$u^*$ of~\eqref{eq:lapla} to define the conductivity
parameter~\cite{Flaherty1972}
\begin{equation}
\sigma = \dfrac{\widehat{R}^4}{\mu}  \displaystyle \int_{A^*} u^* dA^*,
\label{eq:cond}
\end{equation}
so that the expression of the volumetric flux of fluid  
\begin{equation}
Q=\int_A u \, dA=-\dfrac{1}{\mu}\dfrac{dp}{dz}\, \widehat{R}^4
\displaystyle \int_{A^*} u^* dA^* 
\label{eq:flux-def}
\end{equation}
can be re--formulated as the  generalized Ohm's law connecting flux and pressure gradient
\begin{equation}
Q=-\sigma \dfrac{dp}{dz}.
\label{eq:Ohm}
\end{equation}
Notice that the conductivity~$\sigma$ is a function 
of the geometry of the vessel cross section,
this latter being  
itself an unknown of the problem. The coupling with
a structural model for the vessel wall through the pressure loads
closes the problem. 

\medskip

We now go back to the original fluid balance equations~\eqref{eq:cons1} and we integrate Eq.~\eqref{eq:cons1}$_1$ 
on the cross section area. Gathering the resulting equation and the Ohm's law~\eqref{eq:Ohm}, 
we obtain the (equivalent) system: find $Q$ and $p$ such that 
\begin{equation}
\dfrac{dQ}{dz}=0, \qquad Q=-\sigma \dfrac{dp}{dz} \qquad  \qquad  \text{in}~\Omega_f.
\label{eq:fluidmodelQ}
\end{equation}

\begin{remark}
In system~\eqref{eq:fluidmodelQ}, 
the conductivity parameter must be constant in each vessel element. 
For this reason, albeit the blood pressure  in each vessel element is 
a linear function of the~$z$ coordinate, we consider the  
structure to be loaded on the lumen interface with a unique constant pressure~$\overline{p}$ 
function of~$p$
(for example,  its average along the element length) and, thus $\sigma=\sigma(\overline{p})$. 
This approximation is acceptable if the number of elements in each vessels
are chosen in a such a way that the
pressure gradients are not excessively high.
\end{remark}

\subsection{Vessel wall model}
\label{sec:configurations}

\begin{figure}[!hb]
\centering
\includegraphics[scale=1]{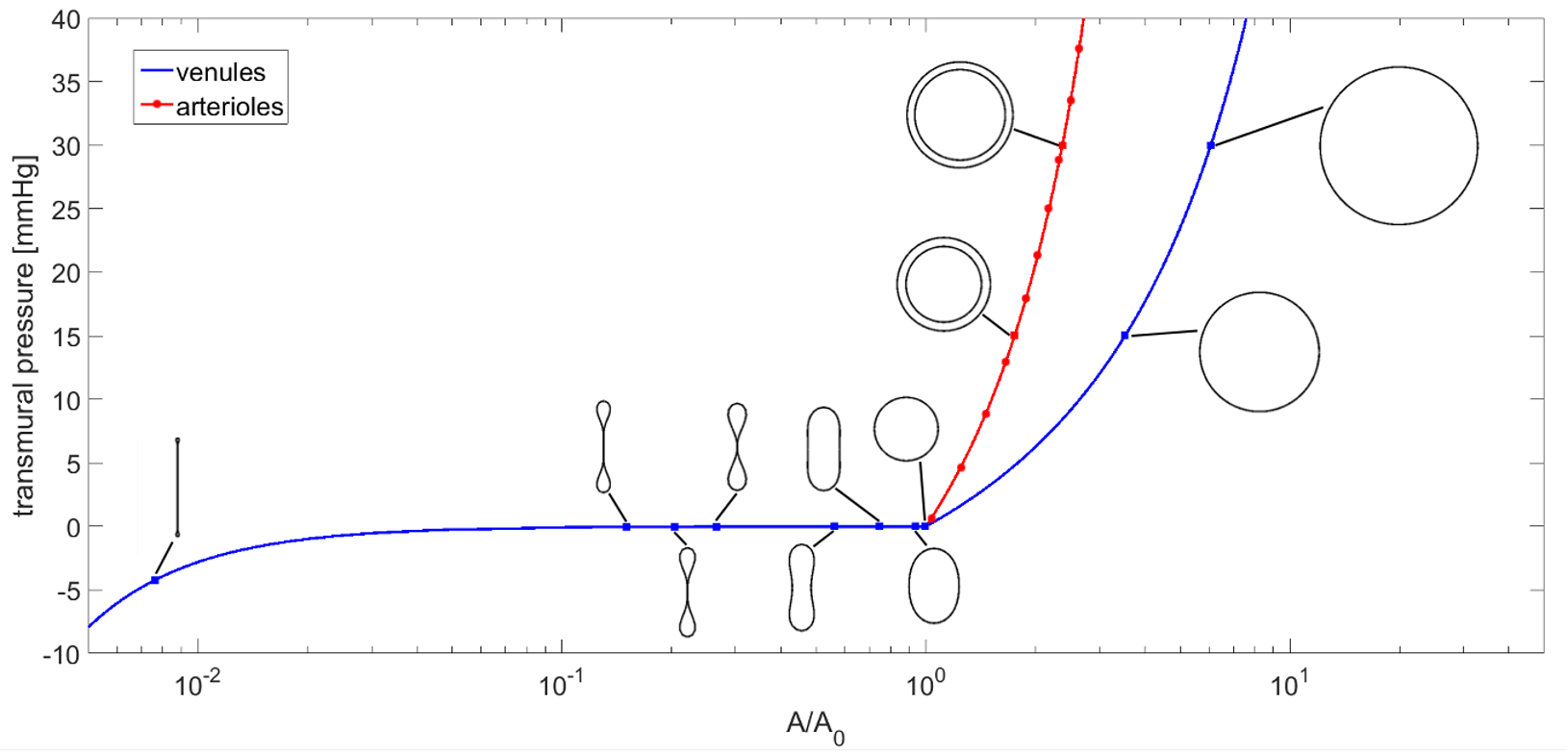}
\caption{Tube laws (transmural pressure vs. area relations) 
 for arterioles and venules as obtained from the present model.   
As customary when representing this curve, the cross section area is normalized over
the cross section area at zero transmural pressure. 
Characteristic  cross sections are sketched for various values of the transmural pressure. 
Observe in proximity of zero transmural pressure 
the presence of a ``snap action'' in venules, {\em i.e.}, a change in
shape over a pressure range so small as to be considered negligible. 
Venules with $A/A_0 <10^{-2}$ are practically collapsed.
The curves are obtained using the same data considered for Fig.~\ref{fig:conductivity}.  
}
\label{fig:tubelaw}
\end{figure}

In order to compute the vessel conductivity from Eq.~\eqref{eq:cond}, we need to dispose of the vessel cross section area and shape
as a function of the pressure loads. In other words, we must build via a structural model 
a tube law, mathematically  connecting
the vessel cross section area  with 
the transmural pressure~$p_{\text{t}}$, defined as the difference
between the luminal pressure~$p$ and the interstitial pressure $p_{\text{e}}$~\cite{Vullo2014}.
We anticipate in Fig.~\ref{fig:tubelaw} the tube law resulting from the present model. Observe, in particular, 
the different behavior of arterioles (thick-walled vessels) 
and venules (thin--walled vessels). Observe also how, for these latter, there exists a physiologically plausible value
of transmural pressure under which the tube is not any more circular but assumes a 
buckled configuration.  
Notice that the cross section does not need to be completely closed for the vessel to be  
``functionally lost to the network''.
As a matter of fact, it suffices the section to be 
small enough to prevent red blood cells passage to compromise its physiological function~\cite{Fry2013}. 

\begin{figure}[hb!]
\centering
\includegraphics[scale=1]{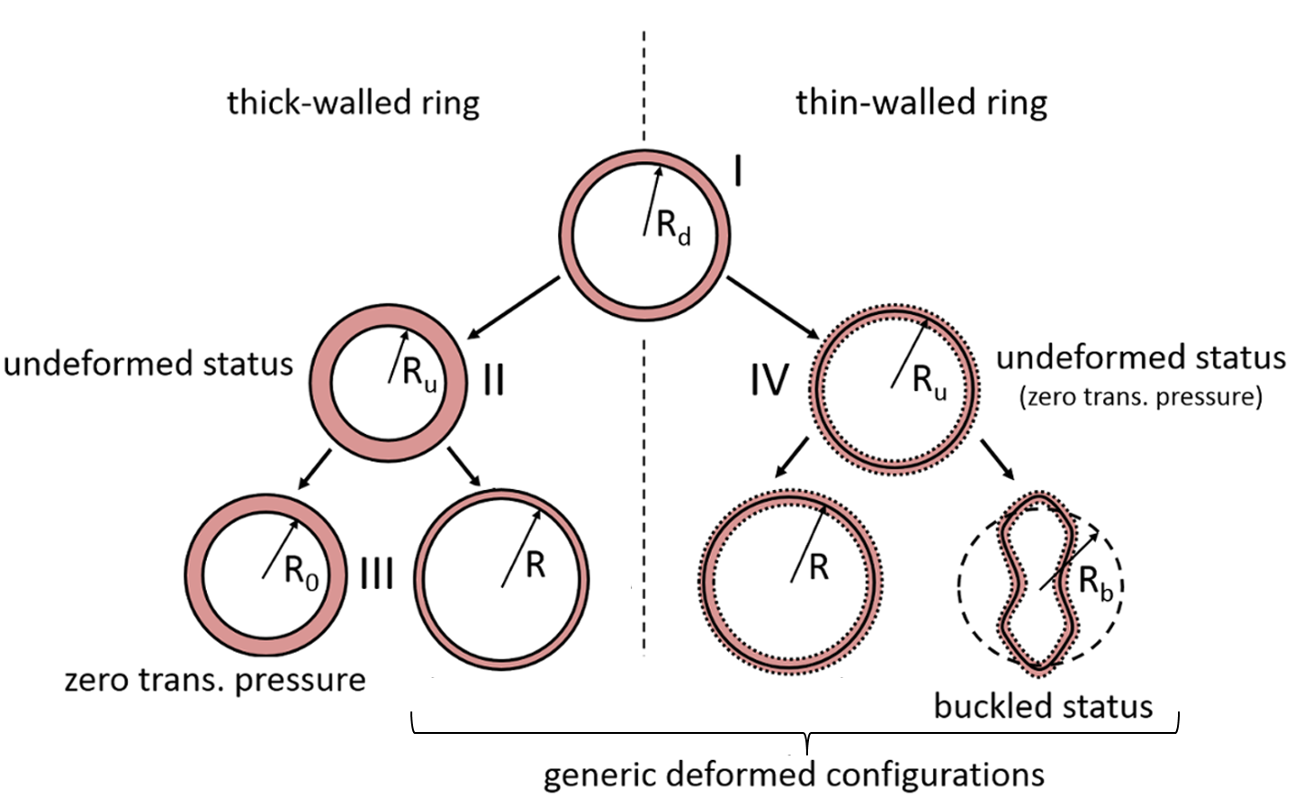}
\caption{
Characteristic configurations in the structural models.
Configuration~I is the experimentally measured ``\emph{in-vivo}'' 
geometry, supposed to be circular. 
The arrows indicate the steps followed in the computations to obtain
a certain configuration from this configuration. 
According to the different modeling chosen as a function of the 
wall thickness-to-radius ratios (see Sect.~\ref{sec:pre-buckl}),
the left side of the figure
refers to thick--walled rings (the internal radius is indicated), while the right 
side to thin--walled rings  (the mean fiber radius is indicated). 
Configurations II and IV are the unloaded configurations, corresponding to 
the stress-free geometry for the thick--walled ring and to the zero transmural pressure
geometry in the thin--walled ring, respectively. 
Notice that in the case of a thick--walled ring the undeformed geometry~II differs from
the zero transmural pressure geometry~III.
We refer to Sect.~\ref{sec:unloaded} for 
a detailed discussion on the computation of the unloaded configuration.
In the last row, we represent generic deformed geometries of the thick and thin--walled ring
cross sections, respectively. In the case of the thin--walled ring, we also consider the possibility of 
section buckling, so that 
the generic deformed cross section is circular if in pre--buckling conditions (left)
or with a non--circular shape if in buckled conditions (right).  
The same terminology
adopted for the radii also applies to the vessel wall thicknesses in the various conditions. 
}
\label{fig:configurations} 
\end{figure}

In the structural model, each vessel element is modeled as an elastic ring
made of elastic (Young modulus~$E$) and incompressible material ($\nu$=0.5), assumed to be 
circular in undeformed conditions (radius~$R_u$, thickness~$h_u$). 
The same cylindrical coordinate system of the fluid model is considered, if not otherwise specified.     
Small deformations are considered, similarly to several works in this field, see {\em e.g.},~\cite{Mikelic2007,Formaggia2009}.
In Fig.~\ref{fig:configurations}, we report the notation required 
for the mathematical discussion and the definition of 
the relevant configurations we will consider.

\subsubsection{Structural model for pre--buckling transmural pressure}
\label{sec:pre-buckl}

On applying the internal and external pressure loads, radial and circumferential 
stresses arise in the ring.  We assume axisymmetry and plane stress conditions.
Let $\eta=\eta(r)$ be the radial displacement of a point of the vessel wall.
\begin{figure}[h]
\centering
\includegraphics[scale=1]{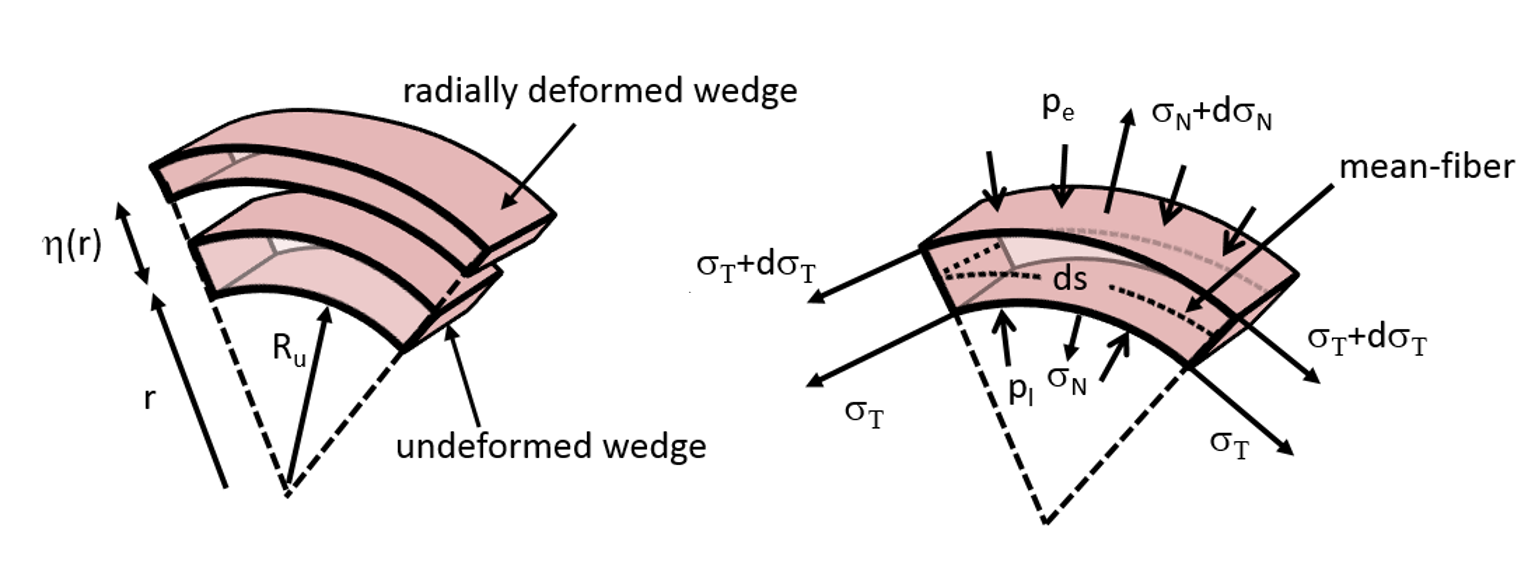}
\caption{Infinitesimal wedge--shaped radial section used for the derivation of the
balance equations for ring model with circular cross section. Left: radial deformation. Right: 
tangential (hoop) stress~$\sigma_{T}$, normal (radial) stress~$\sigma_N$ 
with their increments acting on the wedge faces, along with the pressure loads.}
\label{fig:stress}
\end{figure}
Then, we establish the strain--displacement relations (refer also to Fig.~\ref{fig:stress}) 
\begin{equation}
\label{eq:kinem}
\varepsilon_{N}=\dfrac{d\eta}{dr}, \qquad 
\varepsilon_{T}=\dfrac{\eta}{r},
\end{equation}
$\varepsilon_N$ being the radial strain  and $\varepsilon_T$ the circumferential strain, respectively,
and the pseudo--elastic constitutive equations
\begin{equation}
\label{eq:constitutive}
\sigma_N=\dfrac{E}{1-\nu^2}(\varepsilon_{N} + \nu \varepsilon_{T}),\, \qquad 
\sigma_T=\dfrac{E}{1-\nu^2}(\varepsilon_{T} + \nu \varepsilon_{N}), 
\end{equation}
$\sigma_N$ being the principal radial stress and $\sigma_T$ the principal hoop stress, respectively, and $E=E(p_{\rm t})$
a functional representation of the Young modulus to be discussed at the end of this section.   
We close the problem considering the equilibrium equation
\begin{equation}
\label{eq:equilibrium}
\dfrac{d\sigma_N}{dr}+\dfrac{1}{r}(\sigma_N-\sigma_T)=0, 
\end{equation}
with boundary conditions $\sigma_N(R_u)=-p$ and $\sigma_N(R_{u,e})=-p_e$, with
$R_{u,e}=R_u+h_u$. 
Eq.~\eqref{eq:equilibrium} combined with~\eqref{eq:constitutive} 
and~\eqref{eq:kinem} and the relative boundary conditions gives  
\begin{equation}
\sigma_N=B_1+\dfrac{B_2}{r^2}, \qquad \sigma_T=B_1-\dfrac{B_2}{r^2}, 
\label{eq:stressthick}
\end{equation}
with $B_1=\dfrac{p R_u^{\,\,2}-p_{\text{e}}R_{u,e}^{\,\,2}}{R^{\,\,2}_{u,e}-R^{\,\,2}_{u}}$ and
$B_2=\dfrac{R_{u}^{\,\,2}R_{u,e}^{\,\,2}(p_{\text{e}}-p)}{R^{\,\,2}_{u,e}-R^{\,\,2}_{u}}$.

\begin{figure}[th!]
\centering
\includegraphics[scale=1]{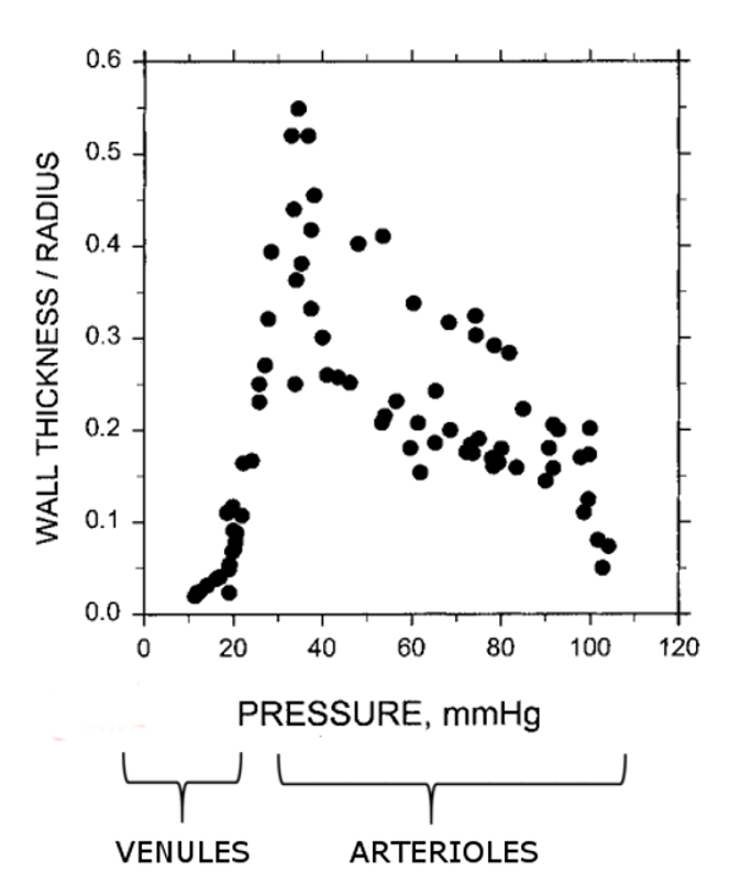}
\caption{
Wall thickness relative to inner vessel radius (parameter $\gamma$ in the model) 
as a function of intravascular pressure (image adapted from~\cite{Pries2001}, data obtained from a meta--analysis 
of literature studies). 
Arterioles and venules are designed to withstand different ranges of luminal pressure.
The arteriolar wall is thus a thick muscularis layer, while the venular wall is a thin
structure. Further, similar, data can be found in~\cite{Lanzer2007,Rhodin1968}.
}
\label{fig:thickness}
\end{figure}

A useful simplification of the expressions in Eq.~\eqref{eq:stressthick} can be obtained for thin--walled structures, since in this case $h_u, h_u^2 \ll R_{m,u}$,  with $R_{m,u} = (R_u+R_{u,e})/2$
(virtual position corresponding to the mean fiber radius). We then obtain the approximations 
\begin{equation}
\sigma_N\simeq 0, \qquad \sigma_T \simeq p_t\dfrac{R_{m,u}}{h_{u}},
\label{eq:stressthin}
\end{equation}
where the second relation represents the well--known Laplace's law. 
Thin--wall  models are considered admissible
till $\gamma \simeq$ 1:10~\cite{Vullo2014}, $\gamma$ being the ratio between the
thickness of ring with respect to the radius.  
As shown in Fig.~\ref{fig:thickness},
the venule wall can thus be considered a thin structure, since $\gamma$ is in 
the range 1:20 to 1:50. Much different is the situation for arterioles, for 
which  $\gamma \simeq 1:3$,
and thus the use of the full expressions in Eq.~\eqref{eq:stressthick} is required.    

In order to derive the expression of the cross section deformed radius,
we combine Eqs.~\eqref{eq:constitutive} with~\eqref{eq:kinem} and
Eq.~\eqref{eq:stressthick} (for thick rings) or Eqs.~\eqref{eq:stressthin}
(for thin rings), respectively, obtaining
\begin{equation}
\left\{
\begin{array}{ll}
R=R_u\left(1+\dfrac{(1-\nu)}{E}B_1-\dfrac{(1+\nu)}{E}\dfrac{B_2}{R_u^2}\right) & \quad \mbox{thick--walled ring}, \\[4mm]
R=R_u\left(1+\dfrac{(1-\nu^2)}{\gamma E} p_{\text{t}}\right) & \quad \mbox{thin--walled ring}, 
\end{array}
\right.
\label{eq:raggio}
\end{equation}
where for thick vessels $R$ denoted the internal radius (blood--vessel interface) while, 
with a slight abuse of notation, for thin vessels $R$ denotes the mean fiber radius.
The deformed ring thickness can be post-computed from incompressibility, yielding
$h=\sqrt{R^2+h_u^2+2h_u R_u}-R$ for thick vessels and $h=h_uR_u/R$ for thin vessels.

\medskip

\emph{Functional representation of the Young modulus.}
Whilst for large blood vessels, especially the carotid, much work has been done, based on 
experimental measures possibly supported by the use of mathematical models 
(see, {\em e.g.},~\cite{Fung2013},\cite{Holzapfel2005}), there is a substantial paucity 
of data and models for the representation of the Young modulus in microvessels.
Given these premises, if one considers for simplicity $E$=const, then 
relations~\eqref{eq:raggio} become linear in the pressure, 
but the corresponding relation transmural pressure vs. cross section area
exhibits two non--physiological features: (i) concave form and  
(ii) absence of saturation at a maximal cross section area for high transmural pressures.
It seems then necessary to use a more complex description than a constant, also in view of the  
different reaction to loads of the components of the vessel wall (collagen, elastin).  
In this work, we use a linear functional dependence of   
Young modulus with respect to  transmural pressure. 
This relation is obtained by fitting data 
from the measurements obtained in~\cite{Zhang2007} by wire
myography in small deactivated arteries and veins of the rat mesenteric circulation.
The computed steepness of the linear relation is such that the Young
modulus passes from a basal value $E_b$ at zero transmural pressure to roughly its double
when the transmural pressure is increased to 50~mmHg. 
In the present example of application of the model to the retinal circulation, 
we set $E_b=0.022$~MPa for arterioles  and $E_b=0.066$~MPa for venules (basal values chosen as in~\cite{Guidoboni2014} for the same microcirculatory district).
Analogous trends can be obtained also considering different sets of measurements, 
for example the ones in~\cite{Holzapfel2005} for human coronary arteries.  
 
\subsubsection{Structural model for buckled thin--walled rings}

\label{sec:thin}
 
\begin{figure}
\centering
\includegraphics[scale=1]{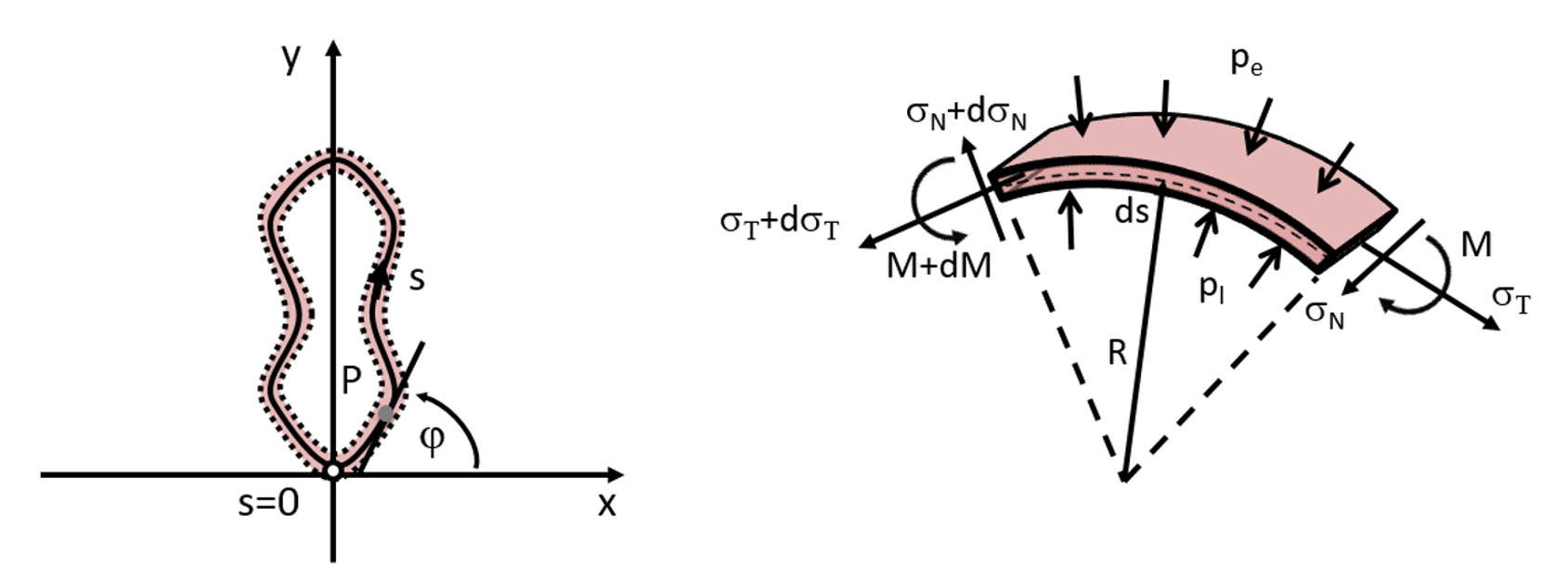}
\caption{Left: coordinate system for 
the thin--walled ring  model in buckled configuration. 
Right: tangential (hoop) stress~$\sigma_{T}$, normal (radial) stress~$\sigma_N$ and  bending
moment~$M$
with their increments acting on the wedge faces, along with the pressure loads.}
\label{fig:buckl}
\end{figure}

When considering the possibility of reaching 
buckled configurations, the model must also keep into account the 
bending actions which actually lead to the loss of axialsymmetry. 
It is convenient in this context to fix a system of Cartesian axes on the bottom 
point of the section (see Fig.~\ref{fig:buckl}, left). We
let~$s$ be the arc--length parameter describing the wall mid-line in counterclockwise direction
from the origin of the axes and we denote by~$\varphi=\varphi(s)$ the 
angle between the positive direction of the~$x$ axis 
and the tangent to the cross section.
The Cartesian coordinates 
 $x=x(s), y=y(s)$ of a point~$P$ identified by arc-length~$s$ are given by 
\begin{equation}
x=\int_0^{s} \cos(\varphi) \, ds, \quad y=\int_0^{s}\sin(\varphi) \, ds.   
\label{eq:coord}
\end{equation}

Fig.~\ref{fig:buckl}(right) shows an element wedge of arch length $ds$ along with
the normal stress~$\sigma_N$, the tangential stress~$\sigma_T$ 
and the bending moment~$M$ arising from the pressure loads.
According to the hypothesis of thin--walled structure, internal actions 
have a constant average value in the radial
direction.
Let $\mathcal{K}(s)=\dfrac{d\varphi}{ds}$ be the local curvature of the section and
$\widehat{\mathcal{K}}=1/R_u$ the curvature of the circular undeformed geometry taken as reference 
configuration.
From the approximate theory of curved beams (see, {\em e.g.},~\cite{Timoshenko1970}), 
the bending moment~$M=M(s)$ has the constitutive form 
\begin{equation}
M=EI\left(\mathcal{K}-\widehat{\mathcal{K}} \right),
\label{eq:momentum}
\end{equation}
where $E I$ is the flexural rigidity, $E$ being the Young modulus (assumed here to be constant and equal to the 
basal value one for total lack of data) and $I={h^3_u}/{12}$ the area moment of 
inertia of the cross section per unit length. 
The balance of bending moments and forces on the infinitesimal wedge--shaped 
radial section of ring per unit axial length is given  by 
\begin{equation}
\dfrac{dM}{ds} = \sigma_N h,  
\qquad \dfrac{d\sigma_N}{ds} = \mathcal{K} \sigma_T - \dfrac{p_{\text{t}}}{h}, \qquad 
\dfrac{d\sigma_T}{ds} = -\mathcal{K} \sigma_N. 
\label{eq:force_mom} 
\end{equation}
Combining Eqs.~\eqref{eq:force_mom} with the Eq.~\eqref{eq:momentum} and using the definition of the
curvature, yields the nonlinear boundary value system 
\begin{equation}
\dfrac{d}{ds}
\left(
\begin{matrix}
\varphi \\[2mm] \mathcal{K} \\[2mm] \sigma_N \\[2mm] \sigma_T
\end{matrix}
\right)
=
\left(
\begin{matrix}
\mathcal{K} \\[2mm] \dfrac{\sigma_N h }{EI} \\[2mm]
\mathcal{K} \sigma_T - \dfrac{p_{\text{t}}}{h} 
\\[2mm] -\mathcal{K}\sigma_N 
\end{matrix}
\right).
\label{eq:BVP_Pavel} 
\end{equation}

Linear stability analysis of system~\eqref{eq:BVP_Pavel} (see, {\em e.g.},~\cite{Tadjbakhsh1967}) 
shows that a buckled non--axisymmetric solution exists for 
every pressure $p_{\text{t}} < p_{\text{t,b}}$, where $p_{\text{t,b}}=-3EI/R_b^3$ is 
the critical transmural pressure corresponding to the lowest energy mode
(azimuthal wavenumber equal to~2). 
When $p_{\text{t}} =  p_{\text{t,b}}$, 
the cross section (of radius~$R_b$ in incipient buckling) loses its circular shape  
due to physical
instability and buckles into an elliptical shape.

For $p_{\text{t}} <  p_{\text{t,b}}$, progressively, the nearest opposite sides  of the section
get close, until they touch if the contact pressure~$p_{\text{t,c}}$ is reached.
The contact point becomes a straight line segment in 
contact if the pressure lowers to the contact line pressure~$p_{\text{t,cl}}$.  As the pressure is further decreased, the length of the contact line increases and the
associated section area tends to zero forming a dumbbell--like shape (see 
the characteristic shapes reported in Fig.~\ref{fig:tubelaw}).

The buckled configurations   
have a two-fold symmetry (since they are related to the
wavenumber~2), which allows for solving system~\eqref{eq:BVP_Pavel} 
just in a fourth of the domain. 
The approach to solve system~\eqref{eq:BVP_Pavel} depends on the value of the transmural pressure, and namely:
\begin{itemize} 
\item[i)] for $p_{\text{t,cl}}<p_{\text{t}}<p_{\text{t,b}}$, 
we compute numerically the solution 
under the  hypothesis of isoperimetrical transformations 
(see also~\cite{Chow2006} for a similar assumption),  by means of the Matlab function 
\texttt{bvp4c} and using the boundary conditions detailed in~\cite{Flaherty1972}.
An appropriate choice of the initial guess shape
is of fundamental importance to kick in the buckling instability in the computation~\cite{Kozlovsky2014}. We 
consider a guess shape with a curvature which is  a small perturbation 
(with parameter $\varepsilon \ll 1$) of the curvature of a circle  
of radius $R_u$, namely
$\mathcal{K}_{\varepsilon}=\dfrac{1}{R_u}\left(1+\varepsilon \cos \dfrac{s}{R_u} \right)$.
This mathematically reproduces the existence in the vessel of imperfections which actually trigger the instability~\cite{Fung2013}.
The value of~$\varepsilon$ must be tuned accordingly to the imposed transmural pressure in order to obtain a physically coherent solution;
\item[ii)]  for  $p_{\text{t}} < p_{\text{t,cl}}$,  the solution 
of~\eqref{eq:BVP_Pavel} can be found from that for $p_{\text{t}}=p_{\text{t,cl}}$ 
by the similarity transformation~\cite{Flaherty1972}
\begin{equation}
\begin{array}{l}
 \varphi(s)=\varphi_{\text{cl}}(s_{\text{cl}}), \quad
\mathcal{K}(s)=(p_{\text{t}}/p_{\text{t,cl}})^{1/3}\mathcal{K}_{\text{cl}}(s_{\text{cl}}), \\[2mm] 
\sigma_N(s)=(p_{\text{t}}/p_{\text{t,cl}})^{2/3}\sigma_{N,\text{cl}}(s_{\text{cl}}), \quad
\sigma_T(s)=(p_{\text{t}}/p_{\text{t,cl}})^{2/3}\sigma_{T,\text{cl}}(s_{\text{cl}}),
\end{array}
\label{eq:trasfo-simil}
\end{equation}
with the coordinate transformation $s=(p_{\text{t,cl}}/p_{\text{t}})^{1/3}s_{\text{cl}}$, where
$0< s_{\text{cl}}< s_1$, $s_1$ being the arc--length of the point of contact
in the configuration corresponding to $p_{\text{t}}=p_{\text{t,cl}}$.
\end{itemize}
Once the solution of system~\eqref{eq:BVP_Pavel} has been computed, 
the non--circular buckled  geometry of the section is reconstructed in Cartesian coordinates from Eqs.~\eqref{eq:coord}.

\medskip

\begin{figure*}[!t]  
 \begin{minipage}[c]{1\textwidth}
 \centering
\includegraphics[scale=1]{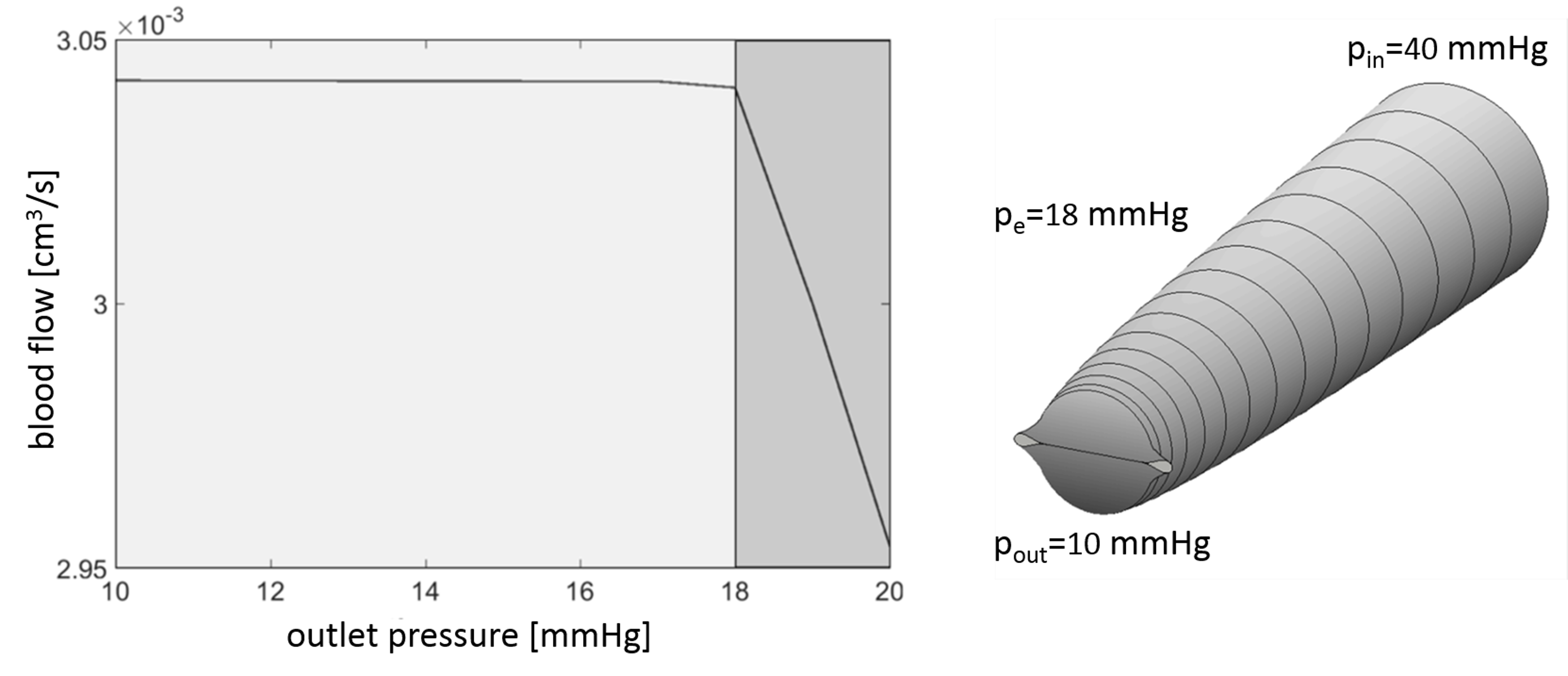}
\caption{Buckling of a single thin--walled distensible vessel.
Left: blood flow in the vessel as a function of~$p_{\text{out}}$. 
As long as $p_{\text{out}}>p_{\text{e}}$ (dark gray region), the flux in the tube
increases  as $p_{\text{out}}$ is decreased. When  $p_{\text{out}} < p_{\text{e}}$ (light gray region),
the downstream portion of the tube enters into buckling instability 
and the flow reaches a plateau, becoming independent of $p_{\text{out}}$. 
Right: 3D configuration
assumed by the tube for $p_{\text out}=10$~mmHg. 
A selected number of cross section shapes (black lines) are highlighted. Notice the very small
dumbbell-shaped cross sections formed at the end of the vessel. 
}
\label{fig:shapes3D}
 \end{minipage}
\end{figure*}

The following case study shows an application of the above described
model and the hemodynamical consequences of vessel buckling.  
We consider a 
 single thin--walled vessel (venule) with inlet pressure
$p_{\text{in}}=40$~mmHg and $p_{\text{e}}=18$~mmHg
and we study the flux for outlet pressure $p_{\text{out}}$ decreasing monotonically 
in the range $[20,10]$~mmHg. 
The vessel has undeformed radius equal to 30~$\mu$m and length equal to 370~$\mu$m.
Simulations are run dividing the vessel into $N_e=800$ consecutive elements,
with progressively smaller elements as the end of the tube is approached. 
In Fig.~\ref{fig:shapes3D}(left), we show the flux
as a function of the outlet pressure. When this latter is decreased, blood flow
increases till $p_{\text{out}}>p_{\text{e}}$. When $p_{\text{out}}=p_{\text{e}}$, 
the downstream portion of the tube enters into buckling.
The flow reaches then a plateau value and it does not depend any more on~$p_{\text{out}}$. 
This trend is in qualitative accordance
with the predictions of the Starling resistor model.
However, in this latter model only two situations are possible, 
fully patent or fully closed vessel cross section.
The distensible behavior simulated in the present work is more complex, since the conductivities  
are consistently coupled with the transmural pressure. 
In Fig.~\ref{fig:shapes3D}(right), we show as an example the 3D~configuration of the 
tube when $p_{\text{out}}=10$~mmHg. Notice the narrow deformed cross sections
in the very downstream portion of the vessel where the low outlet pressure acts. A similar configuration was also 
observed in~\cite{Heil1996}, where a more complex structural shell model coupled
with fluid lubrication theory were used to simulate the experimental setting
of the  Starling resistor device (notice that in~\cite{Heil1996} the upstream and downstream 
cross sections of the tube are maintained fixed). 

\subsection{Computation of the conductivity vs. transmural pressure curve}
\label{sec:riassunto}

The knowledge of the geometry of the  deformed luminal cross section  
as a function of the pressure loads allows for 
computing the element conductivity from relation~\eqref{eq:cond}.
In detail, we proceed as follows:
\begin{itemize}
\item[-] 
if the deformed section remains circular,
relations~\eqref{eq:raggio} explicitly give the radius
of the blood-wall interface. The solution of problem~\eqref{eq:lapla} 
can be then found analytically,  
and yields the usual parabolic Poiseuille velocity profile~\cite{White2006}, 
from which the conductivity can be straightforwardly
computed by integrating~\eqref{eq:cond};

\item[-] if the section is buckled, problem~\eqref{eq:lapla} is numerically solved 
with finite elements on a triangulation of the deformed section.
Vessel conductivity is then obtained by 2D~numerical quadrature of 
the integral~\eqref{eq:cond}. Observe that the buckled configuration is numerically computed
only for a finite number of transmural pressure values. However, a continuous conductivity
curve can be reconstructed by interpolation.
\end{itemize}

Tab.~\ref{tab:cond-calc} summarizes the different expressions/techniques which give the conductivity 
parameter for thick and thin--walled ring elements, respectively.  
\begin{table}[!h]
\centering
{\footnotesize
\begin{tabular}{c|c|c|}
\cline{2-3}
\multicolumn{1}{ c | }{}  &  & \\[-4mm]
 \multirow{2}{*}{}                                                                                  & \multirow{2}{*}{pre--buckling} & \multirow{2}{*}{post--buckling}                                                                                  \\
                                                                                                   &                                &                                                                                                                   \\ \hline

\multicolumn{1}{|c|}{\multirow{3}{*}{\begin{tabular}[c]{@{}c@{}}thick--walled\\ ring\end{tabular}}} & \multirow{3}{*}{$ \sigma(\overline{p})=\dfrac{\pi R_u^4}{8 \mu}\left(1+\dfrac{(1-\nu)}{E}B_1(\overline{p})-\dfrac{(1+\nu)}{E}\dfrac{B_2(\overline{p})}{R_u^2}\right) ^4$} & \multirow{3}{*}{/}                                                                                               \\
\multicolumn{1}{|c|}{}                                                                              &                      &                                                                                                                  \\
\multicolumn{1}{|c|}{}                                                                              &                      &                                                                                                                  \\ \hline
\multicolumn{1}{|c|}{\multirow{3}{*}{\begin{tabular}[c]{@{}c@{}}thin--walled\\ ring\end{tabular}}}  & \multirow{3}{*}{$\sigma(\overline{p})=\dfrac{\pi R_u^4}{8 \mu}\left(1+\dfrac{(1-\nu^2)}{\gamma E} (\overline{p}-p_{\text{e}}) \right) ^4$} & \multirow{3}{*}{\begin{tabular}[c]{@{}c@{}}numerical \\ solution\\ see Sect.~\ref{sec:thin}\end{tabular}} \\
\multicolumn{1}{|c|}{}                                                                              &                      &                                                                                                                  \\
\multicolumn{1}{|c|}{}                                                                              &                      &                                                                                                                  \\ \hline
\end{tabular}
}
\caption{Summary of the different expressions and techniques to obtain the conductivity for thick and thin--walled ring elements. Only positive transmural pressure are considered for the thick--walled rings.  The quantities $B_1$ and $B_2$ are the linear functions of the pressure loads defined in Sect.~\ref{sec:pre-buckl}. Notice
that here we have made explicit the dependence on the pressure indicator $\overline{p}$.}
\label{tab:cond-calc}
\end{table}

\begin{figure}[h!]
\centering
\includegraphics[scale=1]{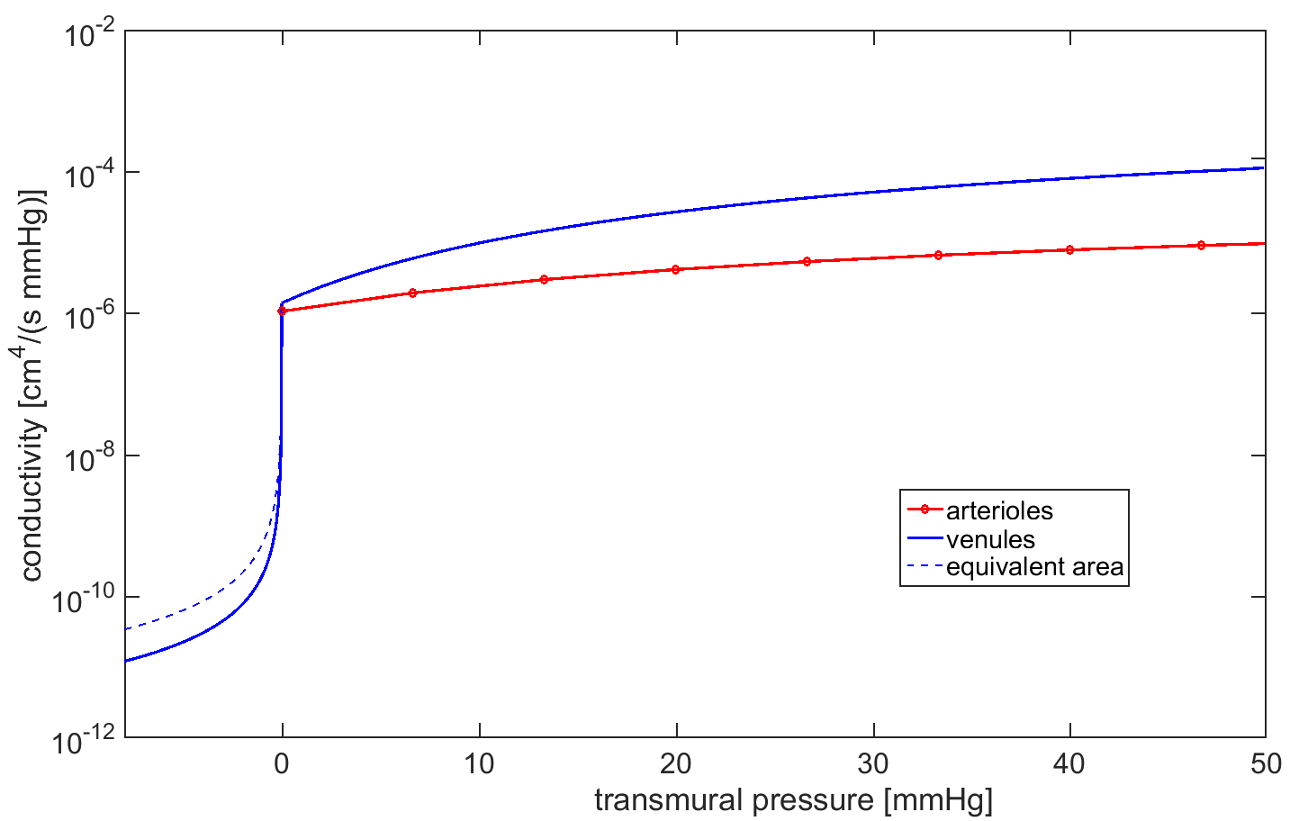}
\caption{Vessel conductivity curve (log scale) plotted against the transmural pressure obtained for a representative 
arteriole (red continuous curve) and venule (blue continuous curve) with radius $R_d=40~\mu$m.
The external pressure is set to $p_{\text{e}}=15$~mmHg, the Young modulus is chosen as discussed in Sect.\ref{sec:phys-par}.
The blue dotted curve represents the conductivity for a circular cross section with the same area of the
buckled configuration at the same value of transmural pressure. Significant discrepancies
arise as $p_{\text{t}}$ decreases in the negative half--plane.
}
\label{fig:conductivity}
\end{figure}

Fig.~\ref{fig:conductivity} depicts an instance of the computed vessel conductivity as a function of
the transmural pressure considering a representative arteriole with $\gamma=0.32$ 
and venule with $\gamma=0.05$, both with $R_u=40~\mu$m.
We choose the Young modulus as discussed in Sect.\ref{sec:pre-buckl} 
and we set~$p_{\text{e}}=15$~mmHg. 
The red curve with circular markers represents the conductivity parameter of the 
arteriole, the continuous blue curve the conductivity of the venule.  
The dashed blue curve in the region of negative 
transmural pressures represents, for comparison, the conductivity of the venule obtained from 
the second relation in Tab.~\ref{tab:cond-calc} (thin--walled ring) 
considering a circular cross section with the same area of the non-circular deformed geometry.

 Notice how this latter curve significantly differs from the 
one obtained with the non--circular geometry, especially in the critical area around the onset of the buckling. 
This motivates us to the explicit computation 
of the buckled geometry.

\subsection{Recovery of  the unloaded configuration}
\label{sec:unloaded}

We conclude the description of the model for a vessel element by dealing with the
problem of recovering the unloaded configuration.
We assume that the 
vessel configurations obtained from  
experimental measurements are circular. Since they do not correspond, 
in general, to unloaded conditions (configuration~II or IV in Fig.~\ref{fig:configurations}), 
we have to solve an inverse problem, 
whose unknowns are the unloaded configuration itself and the stress field
under which the measured deformed configuration is in equilibrium. 
Let 
\begin{equation}
[R_d,h_d]=\mathcal{S}(R_u, h_u; p,p_{\text{e}})
\label{eq:unloaded1}
\end{equation}
be the generic expression of the structural operator, corresponding 
to the thick or thin--walled ring models (direct problem). 
If the undeformed configuration were known, 
the operator~$\mathcal{S}$ would compute the measured geometry 
under given pressure loads. In this context, we have to solve the inverse problem,  
where the unknowns are the unloaded configuration and the stress field
under which the measured deformed configuration is in equilibrium.   
As, in general, the operator~$\mathcal{S}$ cannot be analytically inverted, 
we resort to the fixed--point procedure described in Algorithm~\ref{alg:invprobl}.

\medskip

\begin{algorithm}[!h]
\caption{{\bf : computation of the undeformed geometry}}
\label{alg:invprobl}
%\fbox{\begin{minipage}{0.85\linewidth}
\begin{minipage}{0.85\linewidth}
\vskip 2mm
\begin{algorithmic}
\State {\bf given} $R_\text{d}$, $h_\text{d}$,
$p_{\text{l}},p_{\text{e}}$;
\State {\bf fix} toll, $\omega_r$, $k_{\text{max}}$; 
\State {\bf set} $k$=0, $R_{\text{u}}^{(0)}$=$R_\text{d}$,
$h_{\text{u}}^{(0)}$=$h_\text{d}$;
\While {\,\, \textbf{and}(err $\ge$ toll, $k$ $\le$ $k_\text{max})$  }
\vskip 1mm	
	\State $X^\text{(k)}=\mathcal{S}(R_{\text{u}}^\text{(k)}, 
	h_{\text{u}}^\text{(k)}; p_\text{l}, p_\text{e})$;
	\State $u^\text{(k)}=X^\text{(k)}-R_{\text{u}}^\text{(k)}$;
	\State $ R_{\text{u}}^\text{(k+1)}= \omega_r 
	(R_\text{d}-\text{u}^\text{(k)}) + (1-\omega_r)R_{\text{u}}^\text{(k)}$;
	\State compute $h_{\text{u}}^\text{(k+1)}$ from $R_{\text{u}}^\text{(k+1)}$ using wall incompressibility; 
	\State err$=\| R^\text{(k+1)} - R_{\text{u}}^\text{(k)}\| / \| R_{\text{u}}^\text{(k)}\|$;
	\State $k=k+1$;
\vskip 1mm
\EndWhile
\State $R_{\text{u}}$=$R^{(\text{k})}_{\text{u}}$, $h_{\text{u}}=h^{(\text{k})}_{\text{u}}$
\end{algorithmic}
\end{minipage}
\end{algorithm}

\medskip

Algorithm~\ref{alg:invprobl} is similar to the ones proposed in the computational frameworks
of~\cite{Bols2013,Simonini2015}
in biomedical applications, with the introduction in the present case of a 
relaxation parameter~$\omega_r$. 
We have found in our computations that the number of iterations that are actually needed to converge is related to the
parameter values, being in particular affected by the wall thickness-to-radius ratio and 
by the Young modulus basal value.

\begin{figure}[b!]
\centering
\includegraphics[scale=1]{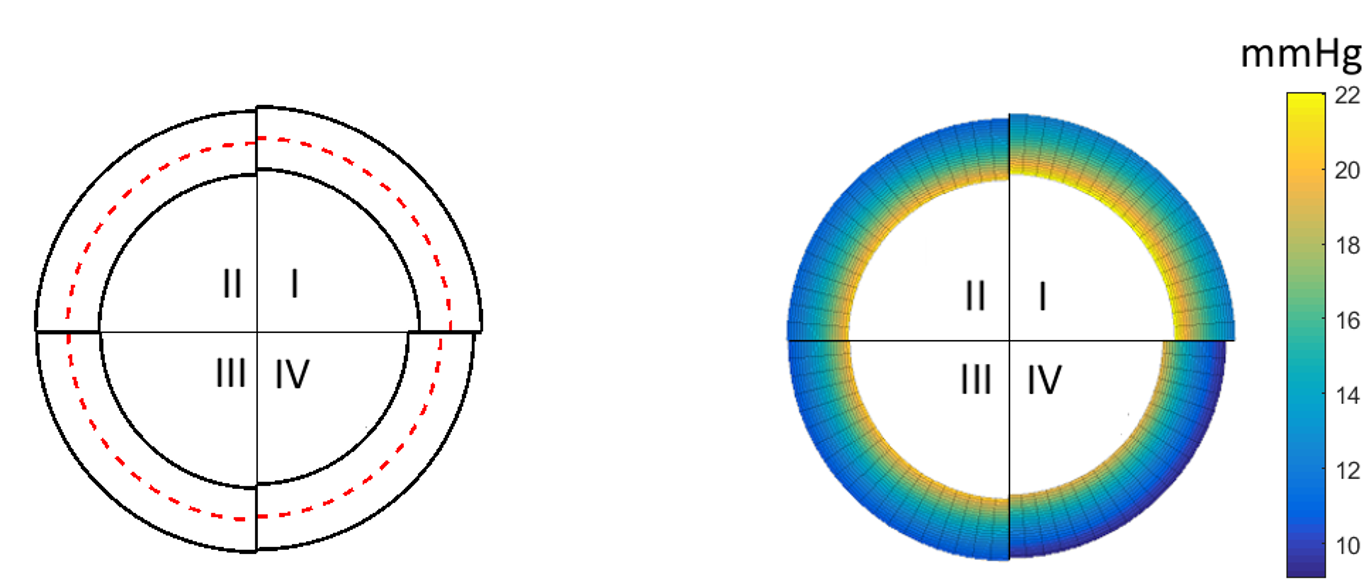}
\caption{Left: configurations I-IV used as unloaded geometries in the 
numerical experiment described in the text.  
The red dotted-line represents the mean radius of the configuration.  
Right: distribution of the hoop stress obtained 
loading configurations I to IV with
$p=20$~mmHg and $p_{\text{e}}=10$~mmHg. 
Stresses color code the corresponding
deformed configurations.}
\label{fig:geo_stress}
\end{figure}

An example of the application of Algorithm~\ref{alg:invprobl} is the following. 
We start from the deformed geometry (configuration~I) of an arteriole 
with circular cross section of  
radius $R_{\text{d}}=40$~$\mu$m, thickness $h_{\text{d}}=12.8$~$\mu$m, 
luminal pressure $p_{\text{d}}=40$~mmHg and external pressure
$p_{{\text{e,d}}}=15$~mmHg (data from~\cite{Takahashi2009}).
The Young modulus is modeled as in Sect.~\ref{sec:pre-buckl}.
From Algorithm~\ref{alg:invprobl}, we obtain the unloaded configuration~II represented in Fig.~\ref{fig:geo_stress}(left).
Convergence till tolerance~$10^{-6}$ is 
achieved after less than~20 iterations with~$\omega_r=0.3$. 
To give an idea of the importance of reconstructing the unloaded configuration,
we also compute 
configuration~III (zero transmural pressure) from~II setting $p=p_{\text{e}}=15$~mmHg
and configuration~IV, which is the unloaded
geometry computed from~I using the thin--walled ring model (see again Fig.~\ref{fig:geo_stress}(left)). 
Observe that configuration~IV is only considered for comparison purposes, since the use of the thin--walled ring model is not appropriate with the present  value~$\gamma=0.32$.

We now apply to configurations I to IV, successively considered as undeformed 
geometries,  
the loads $p=20$~mmHg and $p_{\text{e}}=10$~mmHg.
In Fig.~\ref{fig:geo_stress}(right), we show the resulting deformation and hoop stress fields.

A significant discrepancy in the stress fields  is evident. Configuration~I 
yields stresses which differ of about~1~mmHg with respect to the ones from  
configuration~II (measured vs. unloaded geometry). More significant differences arise
if configuration~IV is used instead of~II (thin vs. thick structure model).
The discrepancies are
an increasing function of the magnitude of the transmural pressure, 
of the individual internal and external pressures, and of the Young modulus (data not reported). 
Tab.~\ref{tab:confi} summarizes the geometrical features of each configuration and
the percentage difference in results with respect to the ones obtained from configuration~I. 
We conclude this section by noting that in the present work we do not consider the existence of pre--stresses (residual--stresses). 
It is well know that, if cut radially, 
vessels spring open releasing the residual stress and approaching the zero-stress state which is a circular 
sector~\cite{Fung2013}. 
This aspect is rather delicate and deserves further future analysis. 

\begin{table}[t!]
\centering
\begin{tabular}{lc|c|c|c|c|}
\cline{3-6}
\multirow{1}{*}{}                        &           & \multicolumn{4}{c|}{Configuration} \\ \cline{1-1}\cline{3-6} 
\multicolumn{1}{|c|}{\multirow{1}{*}{Row}}      &           & I      & II     & III     & IV     \\
\hline
\multicolumn{1}{|c|}{\multirow{1}{*}{1}} & $R$ [$\mu$m]  &  40 & 38.9 & 38.5 & 38.0   \\ \cline{1-6}  
\multicolumn{1}{|c|}{\multirow{1}{*}{2}}  & $\Delta^r R$ [\%]  & / & 2.8 &  3.5 &  5.3   \\ \hline
\multicolumn{1}{|c|}{\multirow{1}{*}{3}} & $h$ [$\mu$m]  & 12.8  & 13.1 & 13.2& 13.5 \\ \cline{1-6} 
\multicolumn{1}{|c|}{\multirow{1}{*}{4}}   & $\Delta^r h$ [\%]  & / &   -2.2&   -3.0&   -5.6 \\ \hline
\multicolumn{1}{|c|}{\multirow{1}{*}{5}}& $\Delta^r \sigma_{T}$ [\%]      &  /  &  9.1 &  11.9 &  19.5\\\hline
\end{tabular}
\caption{Geometrical data and stress values for configurations I to IV and percentage variations with
respect to values of configuration~I. Data refer to the example presented in Sect.~\ref{sec:unloaded}. 
Row 1 and 3: mean radius and wall thickness of 
the initial configurations as in Fig.~\ref{fig:geo_stress}(left); 
rows 2 and 4: percentage variation of the mean radius and wall thickness
in the deformed configuration as in Fig.~\ref{fig:geo_stress}(right); 
row 5: percentage variation of the 
hoop stress~$\sigma_T$ at the mean radius of the deformed configuration as in Fig.~\ref{fig:geo_stress}(right).
The percentage variation is defined as $\Delta^r G:=(G-G_d)/G_d$.}
\label{tab:confi}
\end{table}

\section{Extension to a network of microvessels}
\label{sec:net}

We now consider the study of the fluid field in a complete compliant microcirculatory network, 
organized into incoming arterioles, 
an intermediate capillary bed and draining venules. 
Referring to Fig.~\ref{fig:domain}, 
the network~$\mathcal{T}$ is split  
into~$N_c$ vessels~$\mathcal{V}$, such 
that $\mathcal{T}=\bigcup_{i=1}^{N_c} \mathcal{V}_i$. 
Each vessel~$\mathcal{V}_i$, in turn, is partitioned into $N_e^i$ short consecutive elements~$\mathcal{E}^i$,
such  that $\mathcal{V}_i=\bigcup_{j=1}^{N_e^i} \mathcal{E}^{i,j}$. Notice that
each element in a vessel has its own cross section area
and the number of elements
for each vessel may vary but it has its own cross section area. 
Denoting by $\Omega_f^{i,j}$ the luminal space of the element 
$\mathcal{E}^{i,j}$, the fluid domain is given by
$ \Omega_\mathcal{F}=\bigcup_{i=1}^{N_c} \left(\bigcup_{j=1}^{N_e^i} \,\Omega_f^{i,j}\right)$.   

\medskip

Junctions between vessel elements and different vessels are simply modeled as
nodal points.  
Let $N_{\text{int}}$  be the number of junction nodes. 
For each node $n_k, k=1,\dots, N_{\text{int}}$, we denote by $I_k$ the set of elements which converge in that node. 
Moreover, we denote by $I_k^-$ the subset of $I_k$  
for which~$n_k$
is the first endpoint of the element, {\em i.e.}, an element belongs to $I_k^-$  
if its local axis coordinate is such that $z=0$ in $n_k$. Similarly, we denote by $I_k^+$ the subset of $I_k$  
for which~$n_k$
is the second endpoint of the element, {\em i.e.}, an element belongs to $I_k^+$  
if its local axis coordinate is such that $z=L$ in $n_k$.
At each node $n_k, k=1,\dots, N_{\text{int}}$, we impose continuity of pressure and
conservation of flow (analogue of the electric Kirchhoff's law)
\begin{equation} \label{eq:coup_con1}
\displaystyle \sum_{{i,j} \in I_k^-}-Q^{i,j}+ 
\sum _{{i,j} \in I_k^+} Q^{i,j}=0.
\end{equation}
At the inlet and outlet nodes $n_{in}$ and $n_{out}$ (physiologically, more than
one inlet/outlet can be present in the network, for a total of $N_\text{bdr}$ boundary
nodes), 
we can apply inlet and outlet pressure values (that is, we impose an overall pressure drop, as in the 
simulations presented in this work),
or an inlet flux and an outlet pressure (or viceversa).  

\bigskip \bigskip

\section{Solution procedure}
\label{sec:solproc}

\subsection{Model summary}
\label{sec:model-summ}

The following nonlinear boundary value 
system of PDEs is to be solved in the compliant domain~$\Omega_\mathcal{F}$: \\
given the connectivity of~$\mathcal{T}$, the external pressure, the unloaded
configuration and the mechanical properties of the vessels, 
find the piecewise constant function $Q$ satisfying conditions~\eqref{eq:coup_con1} and 
the continuous--piecewise linear function~$p$ such that in each element it holds 
\begin{equation}
\dfrac{dQ}{dz}=0, \qquad Q=-\sigma(\overline{p})  \dfrac{dp}{dz},
\label{eq:model_summary}
\end{equation}
where the conductivity $\sigma(\overline{p})$ is determined as summarized in Tab.~\ref{tab:cond-calc}.

\subsection{Numerical approximation}
\label{sec:num-approx}

It is convenient to think that 
the discrete counterpart of~\eqref{eq:model_summary}
corresponds to the adoption of a primal mixed finite element. In this framework, $\mathcal{T}$ represents
the \lq\lq triangulation\rq\rq \, of the domain, with elements~$\mathcal{E}$. 
Specifically focusing on the case of prescribed inlet and outlet pressures,
we introduce the finite dimensional spaces (see~\cite{Sacco2014} for a similar procedure, albeit in a different context):
\begin{equation}
\begin{array}{l}
W_{h}: = \{ w_h \in L^2(\Omega_\mathcal{F}): {w_h}_{|_{\Omega_f}} \in \mathbb{P}_0(\Omega_f), \,\forall \,\Omega_f \subset \Omega_\mathcal{F} \}, \\[2mm]  
V_{h; (g_1,g_2)} : = \{ v_h \in C^0(\Omega_\mathcal{F}): 
{v_h}_{|_{\Omega_f}} \in \mathbb{P}_1(\Omega_f),    \,\forall \,\Omega_f \subset \Omega_\mathcal{F}, 
\\[2mm] \qquad \qquad \quad  \,\, v_h(n_{in})=g_1, v_h(n_{out})=g_2 \}. 
\end{array}
\end{equation}

\begin{figure}[h]
\centering
\includegraphics[scale=1.0]{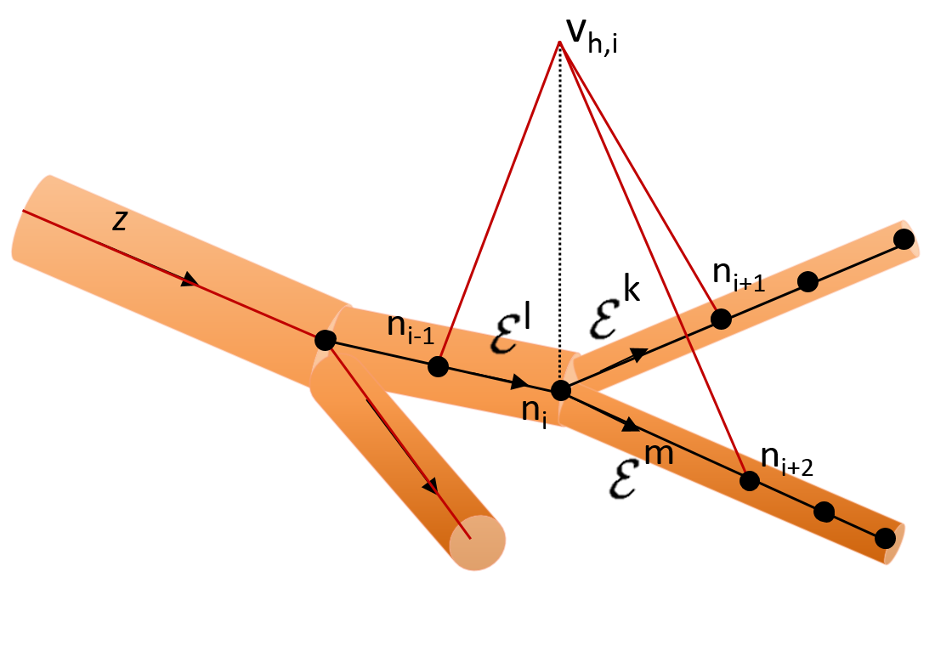}
\caption{Bifurcation of the network including the joining node~$n_i$ and  
the converging vessel elements~$\mathcal{E}^l,\mathcal{E}^k,
\mathcal{E}^m$. On each element, the arrow
indicates the positive direction of the local $z$~axis. This choice implies that
$I_i^+=\{l\},I_i^-=\{k,m\}$. The linear \lq\lq web--like\rq\rq \, shape function~$v_{h,i} \in V_h$
is also represented.}
\label{fig:shape}
\end{figure}

In Fig.~\ref{fig:shape}, we show the shape function $v_{h,i}=v_{h,i}(z) \in V_{h;(g_1,g_2)}$\
relative to node~$i$. This function 
is linear on each element belonging to $I_i^+ \cup I_i^-$ and is such that 
$v_{h,i}=\delta_{i,r}$, $r=1,2,\dots,N_{tot}$, where we have set 
$N_{tot}=(N_{\text{int}}+N_\text{bdr})$. These characteristics
reflect the web--like geometry of the domain.  

We let~$\mathcal{R}(\overline{p})=1/\sigma(\overline{p})$ be the non--negative tube resistance per unit length 
(observe that $\sigma(\overline{p}) > 0$ in the physiological range) 
and
$\forall \, Q_h, w_h \in W_{h}, v_h \in V_{h;(g_1,g_2)}$, $p_h \in V_{h;(p_{in},p_{out})}$, we define the bilinear forms 
\begin{equation}
A(Q_h, w_h; p_h)=\int_{\mathcal{T}} \, \mathcal{R}(\overline{p}_h) \, Q_h \, w_h\, dz,
\quad  
B(v_h,Q_h)=\int_{\mathcal{T}} \,    \dfrac{dv_h}{dz}\,  Q_h \, dz. 
\end{equation}

The numerical solution procedure, including an internal fixed point procedure to solve for the nonlinearities
due to the conductivity, is carried out as described in 
Algorithm~\ref{alg:FEM_discr}. 
%\noindent {\bf Algorithm 2.}\\
\begin{algorithm}[h!]
\caption{{\bf : fixed point iteration to compute the fluid-dynamical 
field on the network}}
\label{alg:FEM_discr}
   \begin{minipage}{0.85\linewidth}
\vskip 2mm
\begin{algorithmic}[h]
\State  {\bf given}  $p_\text{start}$;
\State  {\bf fix}   toll, $\omega_p$, $\omega_Q$, $ k_\text{max}$;
\State  {\bf set}  $p_h^{(0)}=p_\text{start}$, k=0;
\While {\,\, \textbf{and}(err $\ge$ toll, k $\le$ k$_\text{max})$  }
\State   $\overline{p}_h^{(k)}$={ \bf mean}($p_h^{(k)}$) on each vessel;
\State { \bf compute} the cross section geometry from tube law in Fig.~\ref{fig:tubelaw}; 
\State { \bf update} $ \sigma_h^\text{(k)}= \sigma_h(\overline{p}^\text{(k)}); $ 
\State  { \bf solve} 
\State $\qquad$ find $(Q_h^{(k+1)},p_h^{(k+1)}) \in (W_{h}\times V_{h;({p}_{in}, {p}_{out})})$ such that, 
\State $\qquad$  $\forall w_h \in W_{h}, \forall v_h \in V_{h;(0,0)}$
\begin{equation}
\begin{array}{l}
A(Q^{(k+1)}_h, w_h; \overline{p}_h^{(k)})+B(p^{(k+1)}_h,w_h)=0, \\[2mm]
B(v_h,Q^{(k+1)}_h)=0 
\label {eq:sysPMdiscFP}
\end{array}
\end{equation}
	\State $ p_h^\text{(k+1)}= \omega_p 
	p_h^\text{(k+1)} + (1-\omega_p) p_h^\text{(k)}$;
		\State $ Q_h^\text{(k+1)}= \omega_Q 
	Q_h^\text{(k+1)} + (1-\omega_Q) Q_h^\text{(k)}$;
\State err$=\max\{ 
     \| Q_h^\text{(k+1)} -Q_h^\text{(k)}\| / \| Q_h^\text{(k)}\|, \,
     \| p_h^\text{(k+1)} -p_h^\text{(k)}\| / \| p_h^\text{(k)}\|$
     \};
	\State k=k+1;
\vskip 1mm
\EndWhile
\State  {\bf set} $Q_h=Q_h^{(k)},p_h=p_h^{(k)}$.
\end{algorithmic}
\vskip 2mm
\end{minipage}
\end{algorithm}

Referring again to Fig.~\ref{fig:shape} (and omitting for brevity the $k$ superscripts of the
internal iteration procedure), we explicitly 
write the relations obtained from system~\eqref{eq:sysPMdiscFP} in Algorithm~2  
for a bifurcation with joining node~$n_i$, with converging vessel elements~$\mathcal{E}^l,\mathcal{E}^k,
\mathcal{E}^m$ (here, again for brevity, we have used a shortened notation for vessel elements):
\begin{eqnarray}
\mathcal{R}_l(\overline{p}_l)\, Q_{l} L_l+(p_{i}- p_{i-1})  &=&0,
\label{eq:sys1}
\\[2mm]
\mathcal{R}_k(\overline{p}_k)\, Q_{k} L_k+(p_{i+1}- p_{i})  &=&0,
\label{eq:sys2}
\\[2mm]
\mathcal{R}_m(\overline{p}_m)\, Q_{m} L_m+(p_{i+2}- p_{i})  &=&0,
\label{eq:sys3}
\\[2mm]
-Q_l + Q_k+Q_m &=&0
\label{eq:sys4}
\end{eqnarray}

Notice that here we have implicitly made use of the fact
that functions in~$V_{h;(g_1,g_2)}$ are piecewise linear 
continuous over $\mathcal{T}$, so that  they 
ensure the automatic satisfaction of the pressure coupling condition.
The value of the pressure for all the vessels converging in the node~$n_i$ 
is thus uniquely identified by~$p_i$.
Substituting Eq.~\eqref{eq:sys1},\eqref{eq:sys2},\eqref{eq:sys3} (generalized Ohm's laws) in 
Eq.~\eqref{eq:sys4}, we end up  
with a reduced relation in the sole nodal pressure unknowns. Carrying out this procedure for
all the nodes, we obtain the linear algebraic system 
\begin{equation}
\mathcal{M}P=F, 
\label{eq:sistema}
\end{equation}
where
 $P \in \mathbb{R}^{N_{tot} \times 1}$ is the vector of nodal pressure dofs,
$F \in \mathbb{R}^{N_{tot}\times 1}$ is the right--hand side and 
$\mathcal{M} \in \mathbb{R}^{N_{tot}\times N_{tot}}$ is the stiffness matrix.
Boundary conditions are then enforced in system~\eqref{eq:sistema} 
to ensure uniqueness of the solution. In order to achieve convergence in the internal
iteration a relaxation procedure is necessary. 
We have empirically observed that satisfying a convergence criterion on the pressure but also on the fluxes
improves the overall solution. 
To recover fluxes, we use on each element the 
the corresponding Ohm's law.

\section{Numerical simulations}
\label{sec:numres}

\subsection{A practical instance of ``measured'' geometry of a microcirculatory network: the case of eye retina vessels}
\label{sec:geo-net}

The methodology described in the previous sections can address the solution of general unstructured networks. 
However, in this work 
we present simulations performed in arteriolar and venular structures with Y--shaped bifurcations and  with a 
mirrored structural organization for arterioles and venules on each side of the central capillary bed. This choice,
whilst representing an evident idealization of the real anatomy, allows us to carry out a more systematic discussion 
on the results, filtering out the effects of the local irregularity and complexity of the geometry.

In particular, we consider asymmetrically branching networks
which reproduce the microcirculatory district of the eye retina.
Branching is defined according to a modified 
Murray's Law~\cite{Murray1926}:
letting  $D_f$ be the diameter of the larger (father) vessel in a bifurcation and
$D_{d_1}, D_{d_2}$ the diameters of the smaller (daughter) vessels,
the following relation is assumed  
\begin{equation}
D_f^m =D_{d_1}^m + D_{d_2}^m,
\label{eq:Murray}
\end{equation}
where $m=2.85$ is the fractal bifurcation exponent. 
We assume, as described in~\cite{Takahashi2009,Takahashi2014}, that the daughter vessels have (possibly different) diameters, given 
according to 
\begin{equation}
D_{d_1}=c_{d_1/f} D_f, \qquad D_{d_2}=D_f(1-c_{d_1/f}^m)^{1/m}, 
\end{equation}
where $c_{d_1/f}$ is a given proportionality coefficient and where~$D_{d_2}$ has been obtained 
enforcing ~\eqref{eq:Murray}.
The generation of the network is continued as long as the vessel diameter is greater than 6~$\mu$m.  
Observe that $c_{d_1/f} = 2^{-1/m}=0.784$ yields a symmetric dichotomic network with
a constant number of branchings leading from the root to each leaf. 
 The length~$L$ of each vessel  is chosen to be a fractal function of the diameter, 
according to $L = 7.4{D}^{1.15}$, as in~\cite{Takahashi2009}. 
Terminal arterioles and venules are 
connected one-to-one through four parallel capillaries 
with diameter~6~$\mu$m and length~500~$\mu$m~\cite{Takahashi2009}.
Each vessel is then split into equal--sized elements with 
radius and wall thickness equal to that of the vessel itself. 
The undeformed geometry is recovered according to the Algorithm~\ref{alg:invprobl}.
Notice that in this latter configuration the radius of elements of the same vessels
may be different due to nonuniform pressure load.  

To give an idea of the influence of the asymmetry degree of the network, we consider four 
different networks with progressively increasing 
symmetry (that is, with increasing index $c_{d_1/f}$), till
reaching a symmetric dichotomic network $c_{d_1/f}=0.784$.  
In Tab.~\ref{tab:info_gen}, we report 
the values of different features of these networks (total number of vessels, 
min and max route distance of the leaves of the tree,  total cross section and
equivalent resistance of the network in the measured configuration) 
for the considered values of~$c_{d_1/f}$.  
Shown data refer to the arterial side. 
The  trend of the parameters is due to the increasing homogeneity of the network, 
which affects the 
relation between radius and vessel length, and to the constraint of not trespassing the minimum 
diameter. These elements combined together result into 
an equivalent resistance which is more than doubled passing from $c_{d_1/f}=0.5$ to $c_{d_1/f}=0.784$.
 
\begin{table}[ht!]
\footnotesize{
\centering
{\tabulinesep=1.2mm
\begin{tabu}{c|c|c|c|c|c|}
\cline{2-6}
          & \multicolumn{4}{c|}{Asymmetry index~$c_{d_1/f}$} & \multicolumn{1}{c|}{trend} \\
          \cline{1-6}
\multicolumn{1}{|c|}{  Parameter}         & 0.5      & 0.6     & 0.7     & 0.784 & \multicolumn{1}{c|}{ } \\
\hline
 \multicolumn{1}{|c|}{total number of vessels} & 15252& 12415&  9664& 8191 &  $\searrow$ \\ \hline
\multicolumn{1}{|c|}{   min route distance [$\mu$m]} & 1.65 $\cdot 10^{3}$   & 1.84$\cdot 10^{3}$ & 2.36$\cdot 10^{3}$& 3.13$\cdot 10^{3}$& $\nearrow$ \\ \hline
 \multicolumn{1}{|c|}{  max route distance [$\mu$m]}  &1.23$\cdot 10^{4}$ & 7.17$\cdot 10^{3}$  &
 4.49$\cdot 10^{3}$ &   3.13$\cdot 10^{3} $ & $\searrow$   \\ \hline
\multicolumn{1}{|c|}{   total cross section [$\mu$m$^2$] }  &  9.65 $\cdot 10^{5} $ &  7.33$\cdot 10^{5} $ &
  5.85$\cdot 10^{5} $ &  5.17$\cdot 10^{5} $ &  $\searrow$ \\\hline
\multicolumn{1}{|c|}{   eq. resistance [cm$^3$/s/mmHg] }     &  7.84$\cdot 10^{-7} $  &  1.14$\cdot 10^{-6} $
 &  1.43$\cdot 10^{-6} $ &  1.7 $\cdot 10^{-6} $ & $\nearrow$ \\\hline
\end{tabu} }
\caption{Characteristic values of parameters of networks (arterial side only) generated by different
degrees of asymmetry in branching (increasing symmetry moving to the right, 0.784=symmetry). 
The last column indicates the trend of each parameter for increasing symmetry. 
Data correspond to a minimal diameter of 6~$\mu$m,
inlet pressure 40~mmHg, outlet pressure 20~mmHg (values chosen as in~\cite{Takahashi2009}).
}
\label{tab:info_gen}
}
\end{table}

\begin{figure}[!hb]
\centering 
\includegraphics[scale=1]{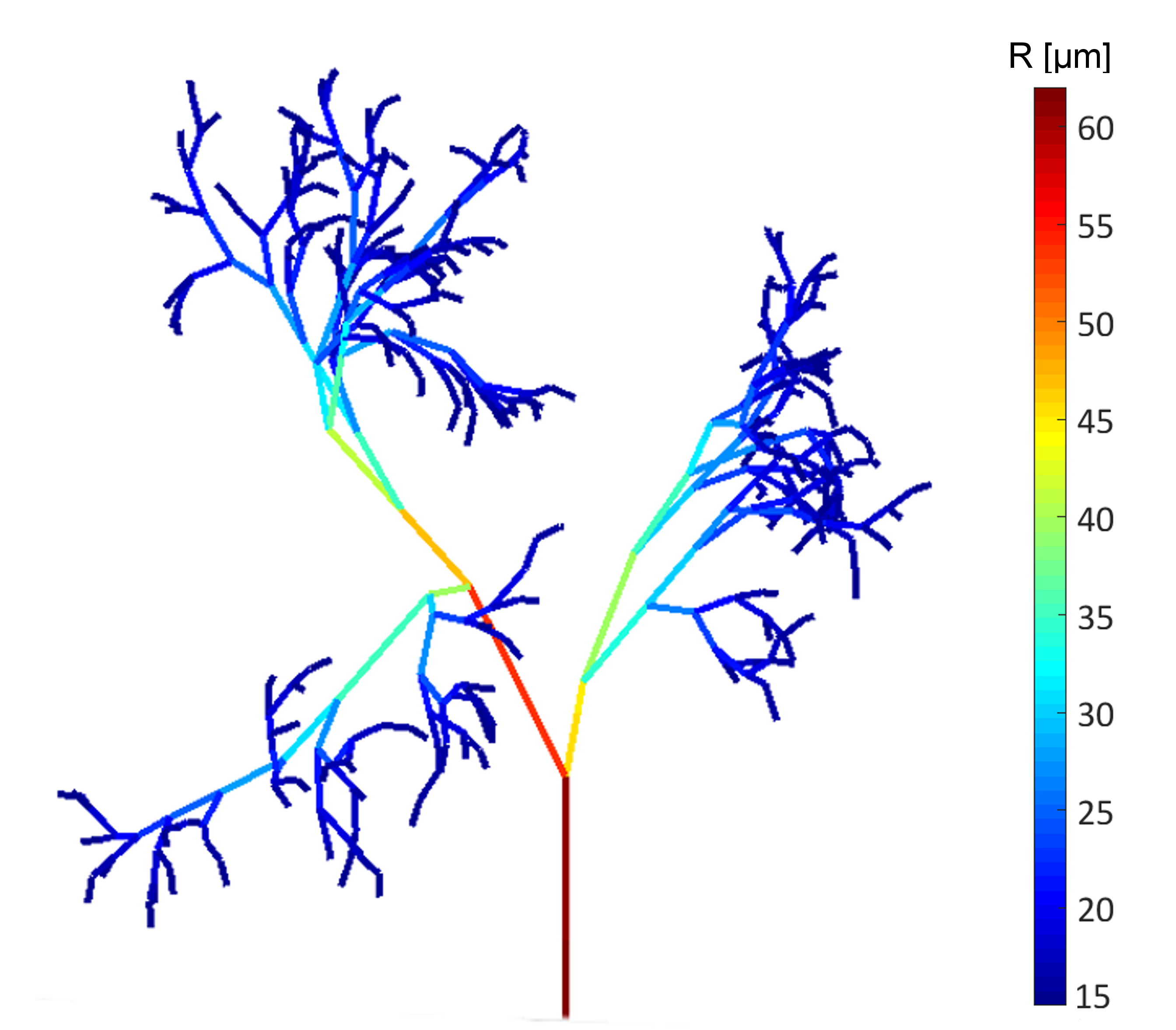}
\caption{Example of a network (arterial side only) considered in the following simulations,
represented in the pseudo--3D space. The network is asymmetric ($c_{d_1/f}=0.7$), as is evident from 
the colors mapping the value of the radii and the different branching structure.  
To improve the readability of the figure, the smallest vessels are not shown.}
\label{fig:network}
\end{figure}

In Fig.~\ref{fig:network} we show an example of network obtained setting~$c_{d_1/f}=0.7$.
Notice that, whilst the above described fractal networks do not possess, {\em per se}, a spatial structure, 
we endow the network of 3D geometrical coordinates by orienting in the space each daughter  branch with
respect to the father with elevation and azimuthal solid angles chosen in a range which respects 
anatomical features. 
This procedure, on the one side, facilitates the visualization of the network and its
physical fields. On the other, more importantly, this allows to locally modify vessel properties
or external conditions in a certain 3D spatial region to reproduce physiological and pathological
alterations of the baseline values.  

\subsection{Physical and numerical parameters}

\label{sec:phys-par}
If not otherwise specified, simulations are run
considering coupled arteriolar and venular networks geometrically described
as in Sect.~\ref{sec:geo-net} on each side with  
$c_{d_1/f}=0.7$. 
The other parameters are chosen as follows. 
We set the network inlet pressure to 42~mmHg and
the outlet pressure to 18~mmHg, respectively. 
The inlet arteriole has radius equal to $62\mu$m, the outlet venule
equal to $72.5~\mu$m. 
We set $\gamma$ 
equal to 0.32 for arterioles and 0.05 for venules.
The first half length of each capillary  is considered to belong to
the arteriolar network, and thus described as a thick--walled ring,
with $\gamma=0.2$. The second  half length is considered to belong to
the venular network, and thus is described as a thin--walled ring,
with $\gamma=0.08$.

Blood viscosity is described according to the model of~\cite{Pries1994}, where 
it is considered to be a function of plasma viscosity ($=1cP$), blood hematocrit ($=45\%$)
and  vessel diameter. 
In this representation, the viscosity decreases as the vessel diameter decreases till 
a diameter of about $40~\mu$m, then it starts to increase again with high steepness
for smaller vessels.  We choose to account for non--circular shapes by using the
concept of hydraulic diameter defined as four times the ratio
between the vessel cross sectional area and the wetted perimeter of the cross-section~\cite{Baskurt2007}.

\medskip

We have studied the influence on the results of the number of element into which
each vessel is partitioned. This difference is extremely small when considering the
physiological range of typical external and internal blood pressures. It is clear
that, were the external pressure be significantly higher than the internal one, 
using a greater number of elements -eventually with an appropriate adaptive refinement - allows to obtain a more accurate description
of the buckling phenomenon. In the following, we set $N_e=3$ in each vessel.

\subsection{Simulation results}
\label{sec:test-cases}

In the following, we present the results obtained  in different test cases which highlight significant network behaviors. 

\medskip

\begin{figure}[!h]
\centering 
\includegraphics[scale=1.0]{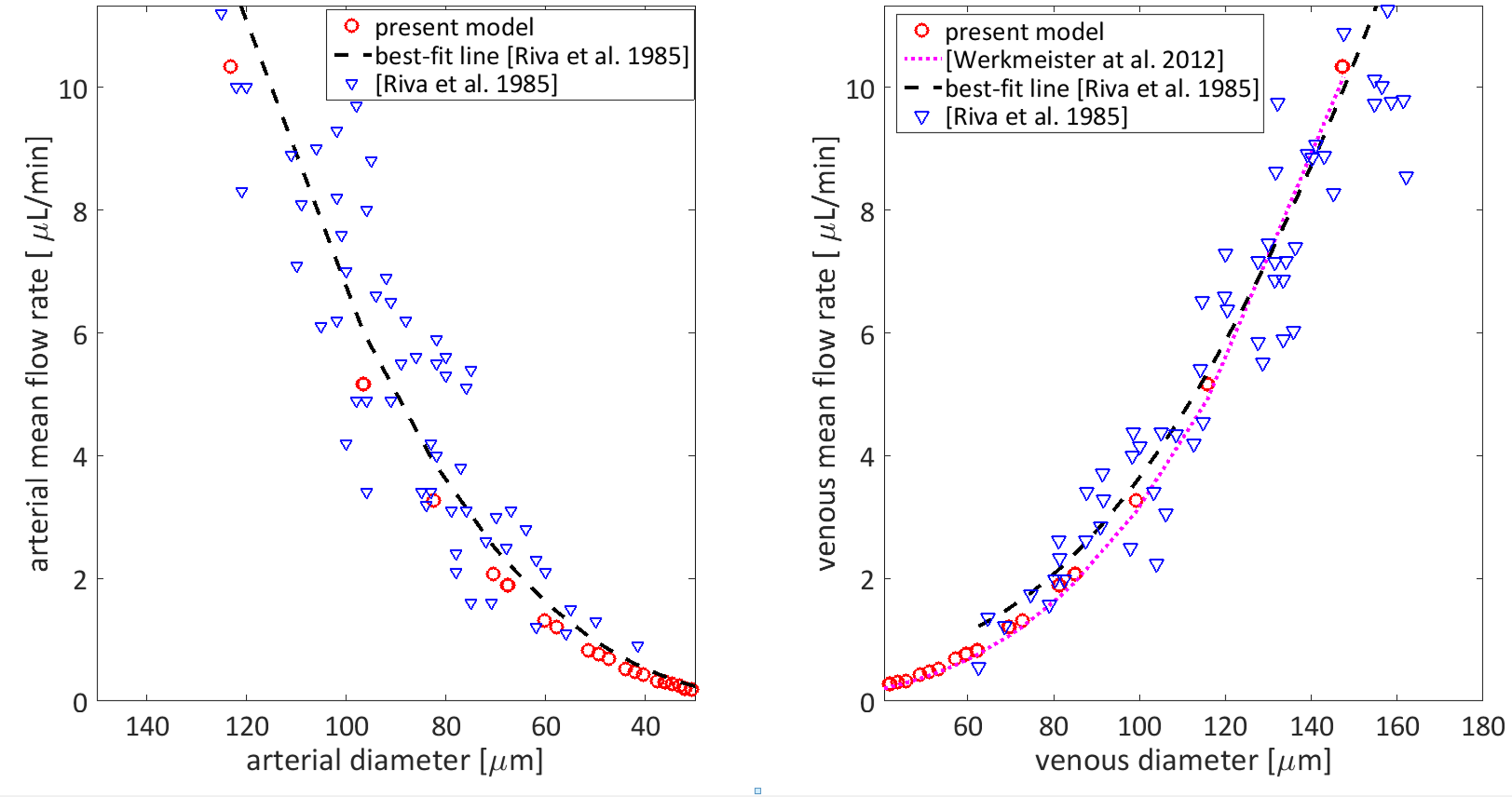}
\caption{Comparison of blood flow computed by the present model on a network generated
by fractal branching and experimental data measured in humans by different authors.}
\label{fig:comparison_Riva}
\end{figure}

\emph{Assessment of the in silico generated networks: comparison with experimental measures.}
To begin with, we assess the fact that the networks we consider can compute physiologically coherent
fluid-dynamical fields. We compare the blood flow we obtain from model simulation considering 
with 
the experimental measures in humans performed by different authors in
the same diameter range. The external pressure is set to $p_{\text{e}}=15$~mmHg,
corresponding to the intraocular pressure of a healthy subject.
The results are shown in Fig.~\ref{fig:comparison_Riva},  which favorably compares volumetric
blood fluxes. Blood velocities (not represented) are also coherent.

\medskip

\begin{figure}[!h]
\centering
\includegraphics[scale=1.]{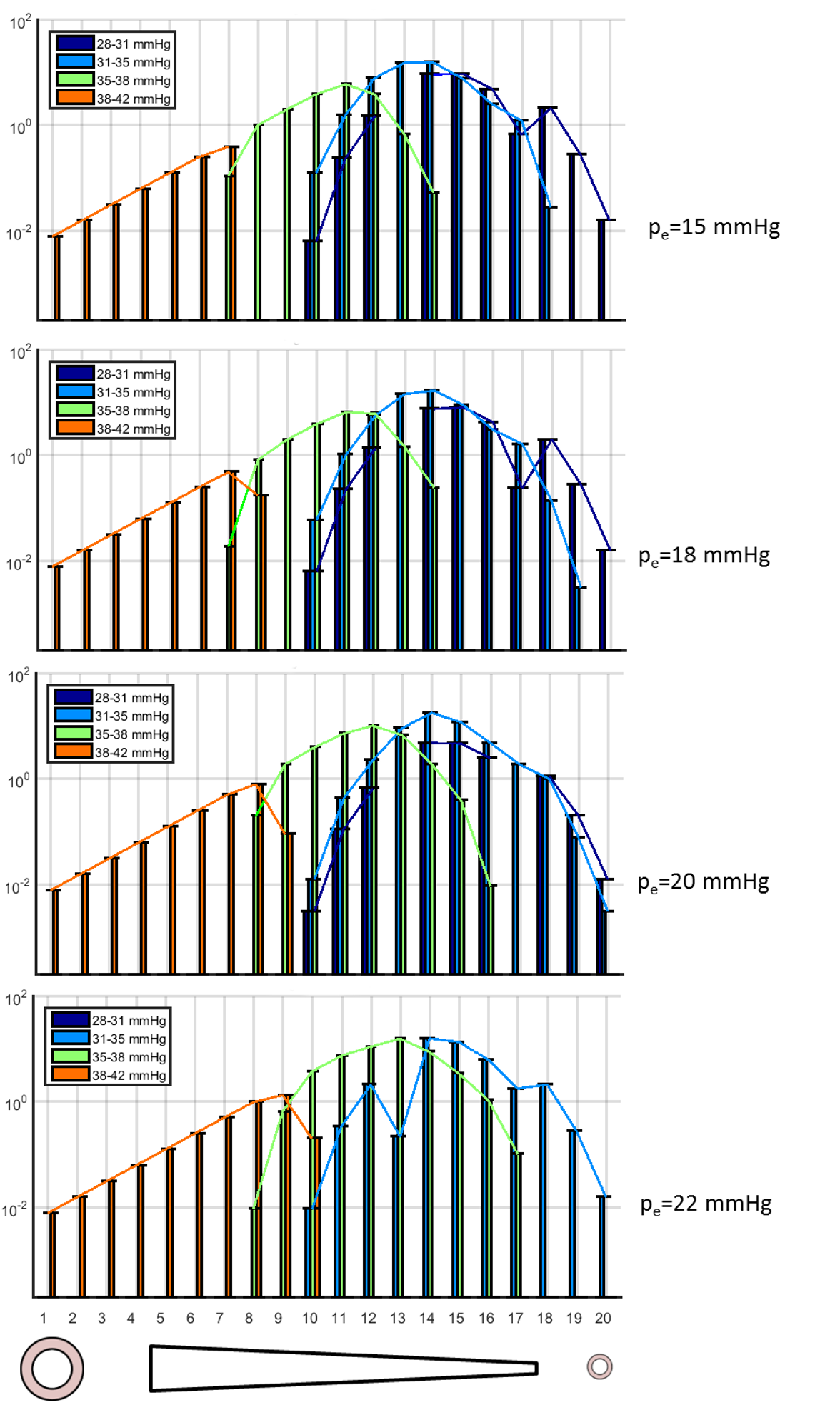}
\caption{
Histogram plots for relative frequencies of blood pressure in the
arterial part of the network binned according to diameter class.
Colored continuous curves link the height of the bars belonging to the 
same pressure range. 
}
\label{fig:pressure-bars-art}
\end{figure}

\begin{figure}[!t]
\centering
\includegraphics[scale=1]{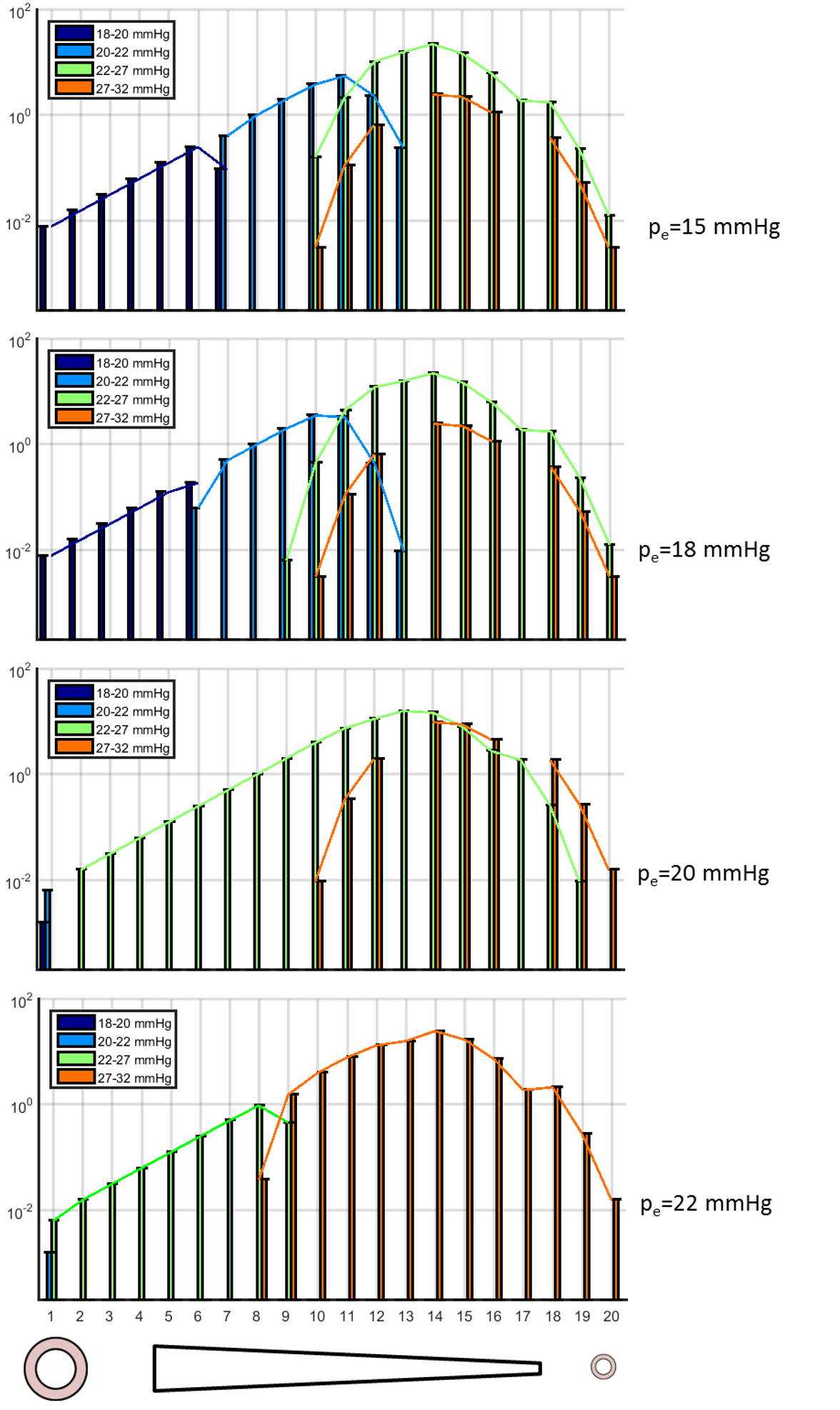}
\caption{
Histogram plots for the relative frequencies of pressures in the
venous part of the network binned according to diameter class.
Colored continuous curves link the height of the bars belonging to the 
same pressure range. 
}
\label{fig:pressure-bars-veins}
\end{figure}

\emph{Pressure field in the network as a function of the interstitial pressure.}
We computationally solve the mathematical model 
choosing a uniform interstitial pressure.
We set successively $p_e=[15,18,20,22]$~mmHg. 
For ease of presentation, we report separately 
the results for arteries and veins, even if the
simulations are run on the coupled model.  
We have binned the vessel into classes according to 
their reference diameters. Class boundaries are obtained by 
partitioning the range between the minimum and maximum diameter
into 20 classes using log10 spacing. 
Class~1 corresponds to larger vessel, 20 to smaller ones. 
For each class, we plot the bars whose heights represent the
relative number (with respect to the total number of arteries and veins, 
respectively) of vessels displaying an average pressure in the 
color--coded range. Colored continuous curves link the height of the bars belonging to the 
same pressure range. 
Fig.~\ref{fig:pressure-bars-art} shows the histogram plots for arteries, 
Fig.~\ref{fig:pressure-bars-veins} for veins.

\medskip

Observe as, for increasing external pressure, peaks for each pressure range shift towards
smaller vessels (larger vessels in the venular network), thus indicating a global increase in blood pressure. 
Lower pressure ranges progressively tend to disappear.
This phenomenon
is to be ascribed to the presence of buckled vessels in veins. 
In the two bottom panels of Fig.~\ref{fig:pressure-bars-veins} we also observe a much more 
sharp compartmentalization of lower pressure ranges in class~1. Again, this phenomenon is to
be ascribed to the presence buckled vessels.
This is also the cause of the larger spreading in pressure ranges attained over the diameter classes
for increasing $p_e$. These differences are more evident in the step $p_e$=20 to
 22 mmHg ($p_e$=18 to 20~mmHg  for the venous network) than in the previous steps. 
Elbow--like turns or ``holes'' in the enveloping
curves are instead due to the asymmetry of the network and the specific bin classification of each vessel.  

\begin{figure}[!h]
\centering
\includegraphics[scale=1]{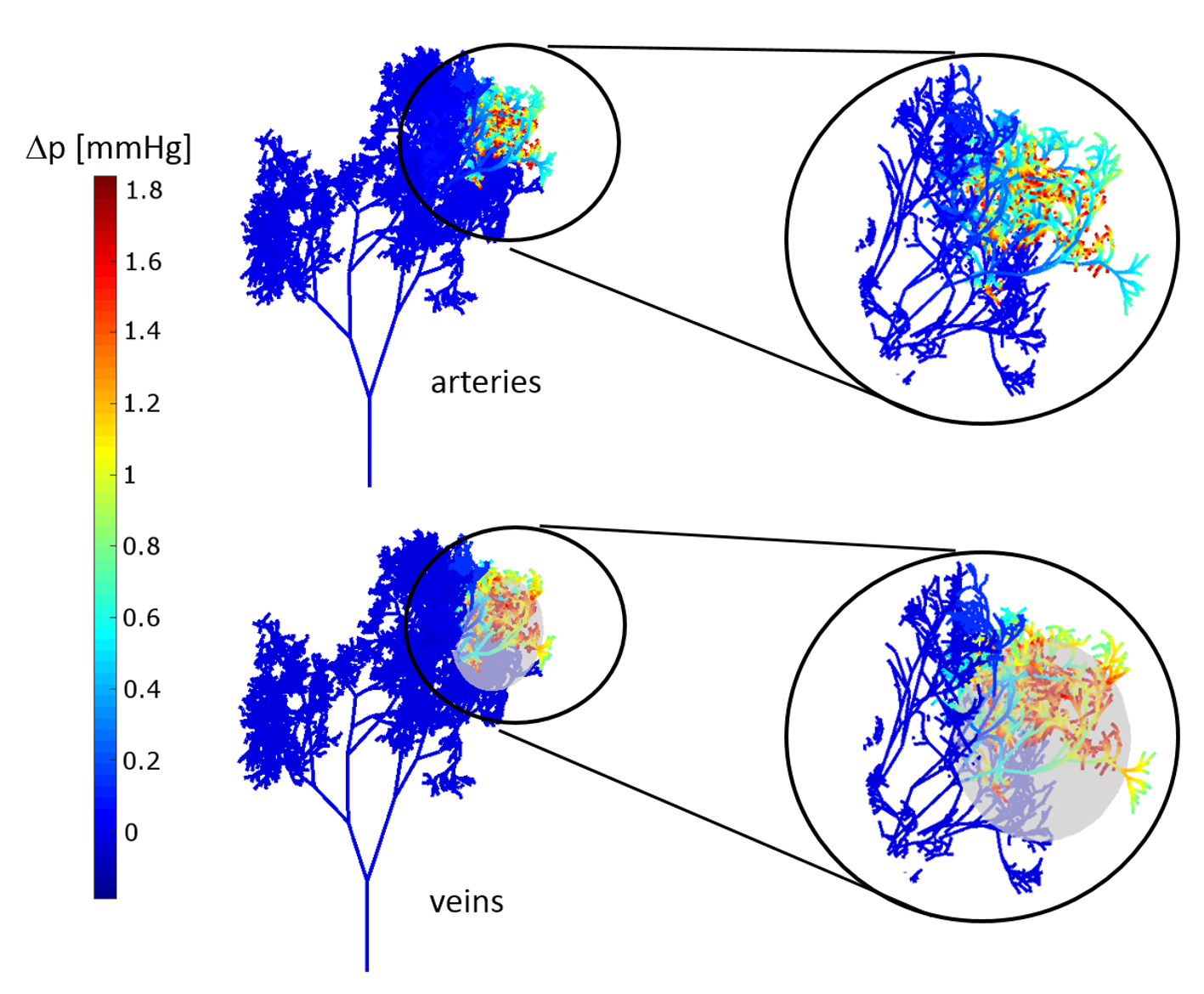}
\caption{Effect of a local increase of external pressure. The gray shaded region
represents the sphere inside which the external pressure is set to $p_e=20$~mmHg, being
elsewhere equal 15~mmHg.
Colors code the net variation $\Delta p$ of blood pressure in 
the arterial and venous networks. Insets on the right part show
zoomed view of the affected network regions.  
}
\label{fig:pall}
\end{figure}

\emph{Network response to local increases of interstitial pressure.}
We artificially increase the external pressure to $p_e=20$~mmHg (elsewhere being equal to 15 mmHg)
in a delimited portion of the network,
located at the interior of a sphere with center in the venous post--capillary zone and radius
chosen to include a sufficient number of vessels (gray--shaded region in Fig.~\ref{fig:pall}(bottom row)).  
This setting simulates, for example, 
the presence of an edema. In Fig.~\ref{fig:pall}, we plot 
color--coded net variation $\Delta p$ of blood pressure in 
the arterial and venous networks along with a zoom of the affected regions.  

In Tab.~\ref{tab:pall-tab} we report the maximal (with sign) percentage variation along the arteriolar and
venular network of
blood pressure, blood flux, cross section area and vessel resistance. Albeit no vessel properly enters into
buckling instability due to the chosen parameters, a strong increase in resistance is observed 
in veins due to the combination of area variation and consequent high rise of blood viscosity, strongly
diameter depended in that region. Observe
also how blood pressure and flow variations are instead similar, this being connected to 
the continuity conditions enforced at the artero-venous interfaces.   
This generates strong flux diversion  and \lq\lq flux stealing\rq\rq\, from other 
vessels~\cite{Causin2016,Pranevicius2002}, localized
in the  four or five generations surrounding the sphere.  Observe that, whilst the 
no arterial vessel is actually included in the sphere, there exists a region of the 
arterial network which is also affected by perturbations.

\begin{table}[ht!]
\footnotesize{
\centering
{\tabulinesep=1.2mm
\begin{tabu}{l|c|c|}
\cline{2-3}
          & \multicolumn{1}{c|}{arterioles} & \multicolumn{1}{c|}{venules}  \\
          \cline{1-3}
 \multicolumn{1}{|l|}{mean blood pressure} & +6.46 \% & +7.10 \%\\ \hline
\multicolumn{1}{|l|}{flow rate } & -15.07 \%   & -15.14 \% \\ \hline
 \multicolumn{1}{|l|}{cross section area}  & -15.20 \%& -55.47 \%  \\ \hline
\multicolumn{1}{|l|}{conductance }  &  -31.84 \% &  -91.64 \%  \\\hline
\end{tabu} }
\caption{Maximal (with sign) percentage variations  of the arteriolar and venular networks in the specified vessel parameters
subsequent to a local increase of external pressure in the spherical region depicted in Fig.~\ref{fig:pall}. 
}
\label{tab:pall-tab}
}
\end{table}

\section{Conclusions}
\label{sec:conclu}

The ability of single body's organs to bring about large selective variations
in the rate of blood perfusion relies on the sophisticated regulatory mechanisms of the
peripheral circulatory districts. Blood flow
regulation is obtained by variation of the vessel diameter, under the effect of both passive
and active actions. In this work, we have focused our attention 
on the first set of mechanisms,
investigating the role of geometrical and structural (the so-called \lq\lq physical\rq\rq) factors in flow
regulation. The key point in this process is compliance: elastic microvessels subject to
mechanical loads undergo deformations of their wall, altering the shape of the domain offered
to blood and, ultimately, resistance to flow. This implies, in turn, a redistribution of the 
flow itself in the network. 

The modeling strategy we have proposed in this work 
uses a simplified description of blood flow and vessel wall-to-flow interaction in order to 
make feasible computations on large networks.
At the same time, important features - peculiar characteristic of these networks- are
retained.  
Blood flow  and pressure drop in each vessel element 
have been connected by a generalized Ohm's law including a conductivity parameter, 
function of the area and shape of the tube cross section. These latter have been determined 
in a consistent way by a thin or 
thick--wall structural model loaded by the internal and external pressure loads.   
A buckling model is considered in the case
of venules, which can experience low/possibly negative values of transmural pressure.

Simulations carried out using the mathematical model show 
that globally increasing the external pressure causes a global increase in luminal pressure.
This process is gradual till buckling occurs somewhere, typically at the outflow, in the venous
district. At this point the process has a sort of discontinuity with a much more marked
pressure increase. 
Locally increasing the external pressure, on the other side,  has an influence which can
be estimated to extend till four or five vessels generations away from the perturbed area.
One important point emerges from the above results. Vessels of the same size may experience  
different intravascular pressure values due to their different location in the network. 
Hence, even though the vessels embedded in a certain tissue can be
classified according to size or branching order, 
the hemodynamical phenomena which are associated with changes in intravascular pressure 
should be analyzed and interpreted with caution, especially 
when considering asymmetrically branching systems.

The present model can be used to investigate alterations in microcirculatory 
beds due to pathological conditions. Hyperglycemia in diabetes mellitus 
is known to cause important structural, biochemical and functional changes 
in the peripheral circulation~\cite{Fowler2008}. 
Alzheimer's disease, atherosclerosis or vascular dementia can 
change vascular stiffness. Elevated systolic pressure (hypertension)  
increases the zero pressure lumen area and   
the buckling pressure, so that an hypertensive vessel
is more likely to collapse than a normal one~\cite{Drzewiecki1997}, with many
possible hemodynamical consequences. 
To address these points in a proper way, the following issues should be taken into
account in the future: 
\begin{itemize} 
\item[i)] inclusion of a more realistic and detailed blood rheology. 
In our work~\cite{Causin2015b}, we have represented blood as a mixture of two 
fluids, plasma and red blood cells, considering the effects of red blood cell partition
at the bifurcations and blood viscosity depending on vessel geometry and variable hematocrit. 
A similar approach could be extended without too many difficulties to the present model; 

 \item[ii)] inclusion of a more realistic model of capillary beds 
and coupling with the surrounding tissue.  
If, on the one side, capillaries show a pretty regular mesh--like organization 
almost ubiquitous in the body, on the other, it is not feasible to face their
complete description. 
Dual mesh techniques as discussed in~\cite{Gould2015} for cerebral microvasculature can be used to study
the dialog between the hemodynamical network and the surrounding tissue.
Alternatively, homogenized models for porous media 
can also be used to effectively face this problem as done in~\cite{Erbertseder2012} for pulmonary alveoli circulation;

\item[iii)] inclusion of \lq\lq acute autoregulation\rq\rq\, capabilities
in arterioles, which actively respond to stimuli to maintain
an adequate blood supply and nutrient delivery to tissue. 
This topic is naturally founded on having a disposal a model 
for blood--tissue interaction as described in point ii).
Possible regulatory laws could be borrowed from the works by 
Ursino and co--authors (see, {\em e.g.},~\cite{Gadda2015} and references therein)
in the case of compartmental models, 
 and from the  
 works by Arciero and co-authors (see, {\em e.g.},~\cite{Carlson2008,Arciero2013}) in the case of a 
simplified representative 1D/segment model of large/small arteries/veins and capillaries. 
In these works,  
the vasoactive
response of the arterioles is modeled via \lq\lq regulatory variables\rq\rq\, which depend on myogenic and metabolic stimuli according to phenomenological laws. 
Similar approaches have been recently adopted
to simulate vessel recruitment by David and co-authors 
(see, {\em e.g.},~\cite{David2015} and references therein)
for realistic network of the cerebral microcirculation,
and by Secomb and co-authors (see, {\em e.g.},~\cite{Fry2013} and references therein).

 \end{itemize}

\bibliographystyle{elsarticle-num} 
\bibliography{biblio}

\begin{thebibliography}{10}
\expandafter\ifx\csname url\endcsname\relax
  \def\url#1{\texttt{#1}}\fi
\expandafter\ifx\csname urlprefix\endcsname\relax\def\urlprefix{URL }\fi
\expandafter\ifx\csname href\endcsname\relax
  \def\href#1#2{#2} \def\path#1{#1}\fi

\bibitem{Pries2003}
A.~Pries, Microcirculation abnormalities: assessment techniques, Medicographia
  25 (2003) 231--236.

\bibitem{microcircbook}
R.~Tuma, W.~Duran, K.~Ley (Eds.), Microcirculation, Elsevier.

\bibitem{Fung2013}
Y.~C. Fung, Biomechanics: {C}irculation, Springer Science \& Business Media,
  2013.

\bibitem{Formaggia2009}
L.~Formaggia, A.~Quarteroni, A.~Veneziani (Eds.), Cardiovascular Mathematics:
  Modeling and Simulation of the Circulatory System, Springer--Verlag, Italy,
  2009.

\bibitem{Brunberg2009}
A.~Brunberg, S.~Heinke, J.~Spillner, R.~Autschbach, D.~Abel, S.~Leonhardt,
  Modeling and simulation of the cardiovascular system: a review of
  applications, methods, and potentials, Biomed. Tech. 54~(5) (2009) 233--244.

\bibitem{Ye1994}
G.~F. Ye, T.~W. Moore, D.~G. Buerk, D.~Jaron, A compartmental model for
  oxygen-carbon dioxide coupled transport in the microcirculation, Ann. Biomed.
  Eng. 22~(5) (1994) 464--479.

\bibitem{Ursino1998}
M.~Ursino, C.~A. Lodi, Interaction among autoregulation, {CO}{$_2$} reactivity,
  and intracranial pressure: a mathematical model, Am. J. Physiol. Heart. Circ.
  Physiol. 274~(5) (1998) H1715--H1728.

\bibitem{Arciero2013}
J.~Arciero, A.~Harris, B.~Siesky, A.~Amireskandari, V.~Gershuny, A.~Pickrell,
  G.~Guidoboni, Theoretical analysis of vascular regulatory mechanisms
  contributing to retinal blood flow autoregulation, Invest. Ophthalmol. Vis.
  Sci. 54~(8) (2013) 5584--5593.

\bibitem{Guidoboni2014}
G.~Guidoboni, A.~Harris, S.~Cassani, J.~Arciero, B.~Siesky, A.~Amireskandari,
  L.~Tobe, P.~Egan, I.~Januleviciene, J.~Park, Intraocular pressure, blood
  pressure and retinal blood flow autoregulation: a mathematical model to
  clarify their relationship and clinical relevance, Invest. Ophthalmol. Vis.
  Sci. (2014) 13.

\bibitem{Roy2012}
T.~K. Roy, A.~R. Pries, T.~W. Secomb, Theoretical comparison of wall-derived
  and erythrocyte-derived mechanisms for metabolic flow regulation in
  heterogeneous microvascular networks, Am. J. Physiol. Heart. Circ. Physiol.
  302~(10) (2012) H1945--H1952.

\bibitem{Dawson2003}
G.~Krenz, C.~Dawson, Flow and pressure distributions in vascular networks
  consisting of distensible vessels, Am. J. Physiol. Heart Circ. Physiol.
  284~(6) (2003) H2192--2203.

\bibitem{Boas2008}
D.~Boas, S.~Jones, A.~Devor, T.~Huppert, A.~Dale, A vascular anatomical network
  model of the spatio-temporal response to brain activation, Neuroimage 40~(3)
  (2008) 1116--1129.

\bibitem{David2009}
T.~David, S.~Alzaidi, H.~Farr, Coupled autoregulation models in the
  cerebro-vasculature, J. Eng. Math. 64~(4) (2009) 403--415.

\bibitem{Causin2015a}
P.~Causin, F.~Malgaroli, A mathematical and computational model of blood flow
  regulation in microvessels: application to the eye retina circulation, J.
  Mech. Med. Biol. 15~(02) (2015) 1540027.

\bibitem{Fry2013}
B.~C. Fry, T.~K. Roy, T.~W. Secomb, Capillary recruitment in a theoretical
  model for blood flow regulation in heterogeneous microvessel networks,
  Physiol. Rep. 1~(3) (2013) e00050.

\bibitem{Lee2004}
J.~Lee, A.~Pullan, N.~Smith, A computational model of microcirculatory network
  structure and transient coronary microcirculation, in: Engineering in
  Medicine and Biology Society, 2004. IEMBS'04. 26th Annual International
  Conference of the IEEE, Vol.~2, 2004, pp. 3808--3811.

\bibitem{Bols2013}
J.~Bols, J.~Degroote, B.~Trachet, B.~Verhegghe, P.~Segers, J.~Vierendeels, A
  computational method to assess the in vivo stresses and unloaded
  configuration of patient-specific blood vessels, J. Comput. Appl. Math. 246
  (2013) 10--17.

\bibitem{Ho2013}
H.~Ho, K.~Mithraratne, P.~Hunter, Numerical simulation of blood flow in an
  anatomically-accurate cerebral venous tree, IEEE Trans. Med. Imaging 32~(1)
  (2013) 85--91.

\bibitem{Kozlovsky2014}
P.~Kozlovsky, U.~Zaretsky, A.~J. Jaffa, D.~Elad, General tube law for
  collapsible thin and thick-wall tubes, J. Biomech. 47~(10) (2014) 2378--2384.

\bibitem{Heil2016}
M.~Heil, A.~L. Hazel, Flow in flexible/collapsible tubes, in: Fluid-Structure
  Interactions in Low-Reynolds-Number Flows, 2016, p. 280.

\bibitem{Aletti2016}
M.~Aletti, J.-F. Gerbeau, D.~Lombardi, A simplified fluid-structure model for
  arterial flow. application to retinal hemodynamics, Comput. Methods in Appl.
  Mech. Eng. 306 (2016) 77–94.

\bibitem{Muller2014b}
L.~M{\"u}ller, E.~Toro, Enhanced global mathematical model for studying
  cerebral venous blood flow, J. Biomech. 47~(13) (2014) 3361.

\bibitem{Ursino1997}
M.~Ursino, C.~A. Lodi, A simple mathematical model of the interaction between
  intracranial pressure and cerebral hemodynamics, J. Appl. Physiol. 82~(4)
  (1997) 1256--1269.

\bibitem{ContarinoSIMAI2016}
C.~Contarino, E.~Toro, A one-dimensional mathematical model for dynamically
  contracting collecting lymphatics: first steps towards a model for the human
  lymphatic network, in: L.~Bonaventura, L.~Formaggia, E.~Miglio, N.~Parolini,
  A.~Scotti, C.~Vergara (Eds.), Proceedings of SIMAI 2016, p. 684.

\bibitem{Canic2006}
S.~{\v{C}}ani{\'c}, C.~Hartley, D.~Rosenstrauch, J.~Tambaca, G.~Guidoboni,
  A.~Mikelic, Blood flow in compliant arteries: An effective viscoelastic
  reduced model, numerics and experimental validation, Ann. Biomed. Eng. 34
  (2006) 575--592.

\bibitem{Sriram2012}
K.~Sriram, B.~Y.~S. V{\'a}zquez, A.~G. Tsai, P.~Cabrales, M.~Intaglietta, D.~M.
  Tartakovsky, Autoregulation and mechanotransduction control the arteriolar
  response to small changes in hematocrit, Am. J. Physiol. Heart. Circ.
  Physiol. 303~(9) (2012) H1096--H1106.

\bibitem{Flaherty1972}
J.~E. Flaherty, J.~B. Keller, S.~Rubinow, Post buckling behavior of elastic
  tubes and rings with opposite sides in contact, SIAM J. Appl. Math. 23~(4)
  (1972) 446--455.

\bibitem{Takahashi2009}
T.~Takahashi, T.~Nagaoka, H.~Panagida, T.~Saitoh, A.~Kamiya, T.~Hein, L.~Kuo,
  A.~Yoshida, A mathematical model for the distribution of hemodynamic
  paramters in the human retinal microvascular network, J. Biorheol.
  23~(77--86) (2009) 2999--3013.

\bibitem{Takahashi2014}
T.~Takahashi, Microcirculation in fractal branching networks, Springer, 2014.

\bibitem{Causin2015b}
P.~Causin, G.~Guidoboni, F.~Malgaroli, R.~Sacco, A.~Harris, Blood flow
  mechanics and oxygen transport and delivery in the retinal microcirculation:
  multiscale mathematical modeling and numerical simulation, Biomech. Model.
  Mechanobiol. (2015) 1--18.

\bibitem{Causin2016}
P.~Causin, F.~Malgaroli, Blood flow repartition in distensible microvascular
  networks: Implication of interstitial and outflow pressure conditions., J.
  Coupled Syst. Multiscale Dyn. 4~(1) (2016) 14--24.

\bibitem{Riva1985}
C.~E. Riva, J.~E. Grunwald, S.~H. Sinclair, B.~Petrig, Blood velocity and
  volumetric flow rate in human retinal vessels, Invest. Ophthalmol. Vis. Sci.
  26~(8) (1985) 1124--1132.

\bibitem{White2006}
F.~White, Viscous fluid flow, McGraw-Hill, 2006.

\bibitem{Vullo2014}
V.~Vullo, Circular Cylinders and Pressure Vessels, Vol.~3, Springer, 2014.

\bibitem{Mikelic2007}
A.~Mikelic, G.~Guidoboni, S.~Canic, Fluid-structure interaction in a
  pre-stressed tube with thick elastic walls {I}: the stationary {S}tokes
  problem, Netw. Heterog. Media 2~(3) (2007) 397.

\bibitem{Pries2001}
A.~R. Pries, B.~Reglin, T.~W. Secomb, Structural adaptation of vascular
  networks role of the pressure response, Hypertension 38~(6) (2001)
  1476--1479.

\bibitem{Lanzer2007}
P.~Lanzer, Mastering endovascular techniques: a guide to excellence, Lippincott
  Williams \& Wilkins, 2007.

\bibitem{Rhodin1968}
J.~A. Rhodin, Ultrastructure of mammalian venous capillaries, venules, and
  small collecting veins, J. Ultrastruct. Res. 25~(5) (1968) 452--500.

\bibitem{Holzapfel2005}
G.~A. Holzapfel, G.~Sommer, C.~T. Gasser, P.~Regitnig, Determination of
  layer-specific mechanical properties of human coronary arteries with
  nonatherosclerotic intimal thickening and related constitutive modeling, Am.
  J. Physiol. Heart. Circ. Physiol. 289~(5) (2005) H2048--H2058.

\bibitem{Zhang2007}
R.~Z. Zhang, A.~A. Gashev, D.~C. Zawieja, M.~J. Davis, Length-tension
  relationships of small arteries, veins, and lymphatics from the rat
  mesenteric microcirculation, Am. J. Physiol. Heart. Circ. Physiol. 292~(4)
  (2007) H1943--H1952.

\bibitem{Timoshenko1970}
S.~Timoshenko, J.~Goodier, Theory of Elasticity (3rd edit.), McGraw-Hill, New
  York, 1970.

\bibitem{Tadjbakhsh1967}
I.~Tadjbakhsh, F.~Odeh, Equilibrium states of elastic rings, J. Math. Anal.
  Appl. 18~(1) (1967) 59--74.

\bibitem{Chow2006}
K.~Chow, C.~Mak, A simple model for the two dimensional blood flow in the
  collapse of veins, J. Math. Biol. 52~(6) (2006) 733--744.

\bibitem{Heil1996}
M.~Heil, T.~Pedley, Large post-buckling deformations of cylindrical shells
  conveying viscous flow, J. Fluid Struct. 10~(6) (1996) 565.

\bibitem{Simonini2015}
I.~Simonini, A.~Pandolfi, Customized finite element modelling of the human
  cornea, PloS {O}ne 10~(6) (2015) e0130426.

\bibitem{Sacco2014}
R.~Sacco, L.~Carichino, C.~de~Falco, M.~Verri, F.~Agostini, T.~Gradinger, A
  multiscale thermo-fluid computational model for a two-phase cooling system,
  Comput. Methods in Appl. Mech. Eng. 282 (2014) 239--268.

\bibitem{Murray1926}
C.~D. Murray, The physiological principle of minimum work i. the vascular
  system and the cost of blood volume, PNAS 12~(3) (1926) 207.

\bibitem{Pries1994}
A.~Pries, T.~Secomb, T.~Gessner, M.~Sperandio, J.~Gross, P.~Gaehtgens,
  Resistance to blood flow in microvessels in vivo, Circ. Res. 75~(5) (1994)
  904--915.

\bibitem{Baskurt2007}
O.~Baskurt, M.~Hardeman, M.~Rampling, H.~Meiselman, Handbook of hemorheology
  and hemodynamics, 1st Edition, Biomedical and health research {V}ol. 69, IOS
  Press, 2007.

\bibitem{Pranevicius2002}
M.~Pranevicius, O.~Pranevicius, Cerebral venous steal: blood flow diversion
  with increased tissue pressure, Neurosurgery 51~(5) (2002) 1267.

\bibitem{Fowler2008}
M.~J. Fowler, Microvascular and macrovascular complications of diabetes,
  Clinical diabetes 26~(2) (2008) 77--82.

\bibitem{Drzewiecki1997}
G.~Drzewiecki, S.~Field, I.~Moubarak, J.~K.-J. Li, Vessel growth and
  collapsible pressure-area relationship, Am. J. Physiol. Heart. Circ. Physiol.
  273~(4) (1997) H2030--H2043.

\bibitem{Gould2015}
I.~G. Gould, A.~A. Linninger, Hematocrit distribution and tissue oxygenation in
  large microcirculatory networks, Microcirculation 22~(1) (2015) 1--18.

\bibitem{Erbertseder2012}
K.~Erbertseder, J.~Reichold, B.~Flemisch, P.~Jenny, R.~Helmig, A coupled
  discrete/continuum model for describing cancer-therapeutic transport in the
  lung, PloS {O}ne 7~(3) (2012) e31966.

\bibitem{Gadda2015}
G.~Gadda, A.~Taibi, F.~Sisini, M.~Gambaccini, P.~Zamboni, M.~Ursino, A new
  hemodynamic model for the study of cerebral venous outflow, Am. J. Physiol. :
  Heart Circ. Physiol. 308~(3) (2015) H217.

\bibitem{Carlson2008}
B.~E. Carlson, J.~C. Arciero, T.~W. Secomb, Theoretical model of blood flow
  autoregulation: roles of myogenic, shear-dependent, and metabolic responses,
  Am. J. Physiol.: Heart Circ. Physiol. 295~(4) (2008) H1572.

\bibitem{David2015}
C.~L. de~Lancea, T.~David, J.~Alastruey, R.~G. Brown, Recruitment pattern in a
  complete cerebral arterial circle, J. Biomech. Eng. 137~(11) (2015) 111004.

\end{thebibliography}

 \end{document}